\documentstyle[12pt,aasms4]{article}

\lefthead{Bekki Kenji \& Yasuhiro Shioya}
\righthead{Stellar populations in gas-rich galaxy mergers}

\begin{document}
\title{Stellar populations in gas-rich galaxy mergers  II.
  Feedback effects of Type Ia and II supernovae}

\author{Kenji Bekki \& Yasuhiro Shioya\altaffilmark{1}} 
\affil{Astronomical Institute, Tohoku University, Sendai, 980-8578, Japan \\
email address: bekki@astroa.astr.tohoku.ac.jp} 

\altaffiltext{1}{Center for Interdisciplinary Research, Tohoku University, 
Sendai, 980-8578, Japan}

\begin{abstract}

We  numerically investigate  chemodynamical 
evolution of major disk-disk galaxy mergers 
in order to explore the origin of mass-dependent chemical, 
photometric, and spectroscopic properties observed in  elliptical galaxies.
We particularly investigate the dependence of the fundamental properties
on merger progenitor disk mass ($M_{\rm d}$). 
Main results obtained in this study are the following five.

(1) More massive (luminous) ellipticals formed by galaxy mergers between more
massive spirals have larger metallicity ($Z$) and thus show redder colors: 
The typical metallicity ranges
from $\sim$ 1.0 solar abundance ($Z\sim 0.02$) for ellipticals
formed by mergers with $M_{\rm d} =  10^{10} M_{\odot}$
to $\sim$ 2.0 solar ($Z\sim 0.04$) for those with $M_{\rm d} =  10^{12} M_{\odot}$.

(2) The absolute magnitude of negative metallicity gradients developed in
galaxy mergers  is more likely to be larger for massive ellipticals.
 Absolute magnitude of metallicity gradient  correlates  with that of
 age gradient in  ellipticals
 in the sense that an elliptical  with steeper negative metallicity gradient
 is more likely to show
 steeper age gradient.

(3) Radial color gradient is  more likely to
be larger for more massive ellipticals,
which reflects that the metallicity gradient is larger for more massive ellipticals.
For example,
typical $U-R$ color gradient  
 ($\Delta U-R / \Delta \log R$)
 for  $0.1 \leq R/R_{\rm e} \leq 1.0$
 is $-0.13$ for ellipticals with $M_{\rm d} =  10^{12} M_{\odot}$
 and $-0.07$ for those with $M_{\rm d} =  10^{10} M_{\odot}$.

(4) Both $\rm Mg_{2}$ line index in the 
central part of ellipticals ($R \leq 0.1 R_{\rm e}$) 
and radial gradient of $\rm Mg_{2}$ ($\Delta \rm Mg_{2} / \it \Delta \log R$) 
are  more likely to be larger for massive ellipticals. 
$\Delta \rm Mg_{2} / \it \Delta \log R$  correlates  reasonably well with
the central $\rm Mg_{2}$ in ellipticals.
For most of the present merger models,
ellipticals show positive radial gradient of $\rm H_{\beta}$ line index.

(5) Both $M/L_{B}$ and $M/L_{K}$, where $M$, $L_{B}$, and $L_{K}$
are total stellar mass of galaxy mergers, $B$-band luminosity, and $K$-band
one, respectively, depend on galactic mass
in such a way that more massive  ellipticals have
larger $M/L_{B}$ and smaller $M/L_{K}$.

The essential  reason for the derived mass-dependent chemical, photometric,
and spectroscopic properties of ellipticals 
is that galactic mass
can largely determine
total amount of metal-enriched interstellar gas,
star formation histories of galaxy mergers, and the effectiveness of Type Ia and
II supernova feedback,
all of which greatly affect chemodynamical evolution of galaxy mergers.

\end{abstract}


\section{Introduction}
 Empirical physical relations in  elliptical galaxies,
 such as the Fundamental Plane
 (FP) and the color-magnitude (CM) relation,
 are generally considered to provide strong constraints 
 on the formation and evolution of elliptical galaxies,
 such as  the formation epoch, the star formation histories,
 and the merging histories (Faber 1973; Visvanathan \& Sandage 1977;
 Djorgovski \& Davis 1987; Dressler et al. 1987;
 Bower, Lucey, \& Ellis 1992;
 Arag$\rm \acute{o}$n-Salamanca et al. 1993; 
 Djorgovski, Pahre, \& de Carvalho  1996; Franx \& van Dokkum 1996).
 Physical origins of these relations accordingly
 have been extensively explored  by observational and
 theoretical studies from variously different points of views
 (e.g., Arimoto \& Yoshii 1987; Bender, Burstein, \& Faber 1992;
 Djorgovski et al. 1996; Kauffmann \& Charlot 1997).
 Among possibly key determinants for the origins,
 the nature of stellar populations of ellipticals,
 such as age and metallicity distribution, is  recognized as 
 a primarily important factor,
 principally because stellar populations are generally
 considered to affect both the slope and tightness of these
 empirical  relations.
 Origins of such important nature of stellar populations in elliptical galaxies have
 been indeed investigated mostly in the context of
 monolithic collapse model with the so-called
 galactic wind (Larson 1975; Carlberg 1984; Arimoto \& Yoshii 1987;
 Bressan et al. 1996, Gibson and Matteucci 1997),
 they 
 have not been so extensively explored,
 however, in the context of the merger scenario that
 ellipticals are formed by major disk-disk galaxy mergers
 (e.g., Toomre \& Toomre 1972).
 Accordingly,
 it is vital to investigate whether or not 
 the merger scenario,  which has been already confirmed
 to have $some$ advantages in
 reproducing a number of important structural and kinematical
 properties of ellipticals (e.g., Barnes \& Hernquist 1992),
 can also reproduce fundamental properties
 of stellar populations inferred from empirical relations of
 elliptical galaxies.

 There are only a few studies trying to clarify  the origin
 of  stellar populations of elliptical galaxies along the merger scenario.
 Schweizer \& Seitzer (1992) demonstrated  that the bluer integrated $UBV$ color
 of elliptical galaxies with morphologically fine structure can be explained by
 the secondary starburst induced by major disk-disk galaxy mergers.
 Fritze -v. Alvensleben \& Gerhard (1994) showed that even late galaxy
 mergers with their age $\sim$ 12 Gyr can develop typical colors
 of ellipticals within 3 $\sim$ 4 Gyr after the secondary starburst.
 Kauffmann \& Charlot (1997) constructed   a semi-analytic model of
 elliptical galaxy formation, which is based upon
 the hierarchical clustering in CDM universe and includes  rather simple
 chemical enrichment  process, and thereby
 demonstrated  that the origin of the color-magnitude relation of elliptical
 galaxies can be reproduced successfully  in the CDM model of galaxy formation.
 These previous studies, which are based on  $one$-$zone$ chemical evolution
 models, have completely neglected the importance of dynamical effects on
 chemophotometric  evolution of galaxy mergers.
 Since star formation history, which is one of the most critical determinants
 for fundamental properties of stellar populations,
 is considered to depend largely  on galactic dynamics in
 various ways (e.g., Kennicutt 1989; Larson 1992),
 we should investigate dynamical effects on stellar populations 
 of galaxies in an admittedly self-consistent manner,
 in order to understand more deeply and thoroughly 
 the nature of stellar populations of elliptical galaxies.

 The purpose of this paper is to investigate how dynamical 
 effects of galaxy merging, in particular,
 violent relaxation combined with gaseous dissipation,
 affect global properties of stellar populations (e.g.
 mean stellar metallicity and age) and their radial
 gradients.
 The present study is an extended version of our previous
 study (Bekki \& Shioya 1998, hereafter referred to as Paper I),
 which is the first step toward the understanding 
 of a close physical relationship
 between dynamical evolution and chemical one
 in high redshift galaxy mergers (the redshift of $z = 1 \sim 2$)
 between  disk galaxies with the gas mass fraction
 larger than 0.2 that is
 a typical value of the present late-type spirals.
 The importance of investigating high redshift 
 gas-rich galaxy mergers is
 described in Paper I.
 The present chemodynamical model 
 is more realistic and sophisticated than our previous one
 in the following three points:
 (1) Instead of using instantaneous chemical mixing approximation,
 we consider time-delay between star formation
 and metal ejection mainly from Type Ia and II supernovae.
 (2) Feedback effects of Type Ia and II supernovae 
 on dynamical and chemical evolution of galaxy mergers
 are incorporated.
 (3) Time-evolution of  enrichment processes 
 is  solved for each of chemical components, 
 H, He, Mg, O, and Fe.
 This chemodynamical model enables us to reveal not only
 physical origins of global (mean) properties of
 stellar populations, which are the key
 determinants for galactic empirical physical  relations,
 but also radial gradients of stellar populations
 in elliptical galaxies.
 By using the present chemodynamical model,
 we  particularly investigate
 the origin of $mass$-$dependent$ chemical, photometric,
and spectroscopic properties observed in  elliptical galaxies. 
The layout of this paper is as follows.
In \S 2, we summarize  numerical models used in the
present study and describe   methods for
analyzing the stellar populations produced by dissipative galaxy
mergers with star formation.
In \S 3, we demonstrate how  a number of fundamental  characteristics 
of stellar populations in merger remnants 
depend on galactic mass.  
In \S 4, we provide a number of implications on the formation and evolution
of elliptical galaxies, based on the derived results.
The conclusions of the present study are given in \S 5.

\section{Model}
Chemical, photometric, and spectroscopic 
evolution of galaxies are basically
determined by age and metallicity distribution of stellar 
populations within galaxies.
Fundamental properties of the distribution are furthermore
controlled by star formation histories,
which are considered to  depend strongly on dynamical and kinematical properties
of galaxies such as local dynamical instability (e.g. Kennicutt 1989;
Larson 1992).
Hence we numerically solve the time evolution of chemical, photometric,
and spectroscopic 
evolution of galaxy mergers, based on
the dynamical evolution of galaxies. 
Firstly we describe a  numerical model for dynamical evolution of
galaxy mergers,
including structure and kinematics of merger
progenitor disks, initial orbital configurations  of galaxy  mergers,
and the prescriptions of dissipative process (\S 2.1).
Secondly, we describe a star formation model, in which
gas consumption rate of star formation and the effectiveness of Type Ia and II supernovae
feedback are given in detail (\S 2.2).
Thirdly we give the method for analyzing the chemical enrichment process
during dissipative galaxy merging with star formation 
and the photometric properties of merger  remnants (\S 2.3).
Lastly, we describe the main points of analysis of the present study (\S 2.4). 

\subsection{Merger  model}

\subsubsection{Structure and kinematics of merger progenitor disks}

 We construct  models of galaxy mergers between gas-rich 
 disk galaxies with equal mass by using Fall-Efstathiou model (1980).
 The total mass and the size of a progenitor disk are $M_{\rm d}$
 and $R_{\rm d}$, respectively. 
 From now on, all the mass and length are measured in units of
  $M_{\rm d}$ and  $R_{\rm d}$, respectively, unless specified. 
  Velocity and time are 
  measured in units of $v$ = $ (GM_{\rm d}/R_{\rm d})^{1/2}$ and
  $t_{\rm dyn}$ = $(R_{\rm d}^{3}/GM_{\rm d})^{1/2}$, respectively,
  where $G$ is the gravitational constant and assumed to be 1.0.
  Dimensional values for these  units in each model are given later.
  In the present model, the rotation curve becomes nearly flat
  at  0.35  $R_{\rm d}$ with the maximum rotational velocity $v_{\rm m}$ = 1.8.
  The corresponding total mass $M_{\rm t}$ and halo mass $M_{\rm h}$
  are 3.8 and 2.8 in our units, respectively.
  The radial ($R$) and vertical ($Z$) density profile 
  of a  disk are  assumed to be
  proportional to $\exp (-R/R_{0}) $ with scale length $R_{0}$ = 0.2
  and to  ${\rm sech}^2 (Z/Z_{0})$ with scale length $Z_{0}$ = 0.04
  in our units,
  respectively.
  The cut off radius for
  mass density  of the halo component is   1.2 in our units.
  Velocity dispersion of the halo component at a given point
  is set to be isotropic and given
  according to the  virial theorem.
  In addition to the rotational velocity made by the gravitational
  field of disk and halo component, the initial radial and azimuthal velocity
  dispersion are given to disk component according
  to the epicyclic theory with Toomre's parameter (\cite{bt87}) $Q$ = 1.0.
  This adopted value for $Q$ parameter
is appreciably smaller compared with the value required 
 for stabilizing the initial disk
against the non-axisymmetric dynamical instability (e.g. bar instability).
The reason for this adoption is that the initial disk is assumed to
be composed mostly of interstellar gas and thus random kinetic energy
in the disk is considered to be rather smaller because of gaseous
dissipation in  the disk. 
  The vertical velocity dispersion at given radius 
  is  set to be 0.5 times as large as
  the radial velocity dispersion at that point, 
  as is consistent  with 
  the observed trend  of the Milky Way (e.g., Wielen 1977).

\subsubsection{Gas dynamics}

  The collisional and dissipative nature 
  of the interstellar medium (ISM) is  modeled by the sticky particle method
  (\cite{sch81}).
It should be emphasized here that this discrete cloud model can at best represent
the $real$ interstellar medium of galaxies  in a schematic way. 
 Actually, considerably complicated nature of
interstellar medium, which can be modeled as
three phase structure (`hot', `warm', and `cool' gas) by
 McKee \& Ostriker (1977),
would not be
so simply modeled by the `sticky
particle' method in which gaseous dissipation is modeled by ad hoc
cloud-cloud collision: Any existing numerical method  could
not model the $real$ interstellar medium in an admittedly proper
way. 
More elaborated numerical modeling for real interstellar medium
would be  necessary for 
our further understanding of dynamical evolution 
in dissipative galaxy mergers. 
  To mimic the
  galaxy mergers which are 
occurred at   high redshift and thus very dissipative
because of a  considerably larger amount of  interstellar gas in the 
progenitor disks,
  we assume that the fraction of gas mass in
  a disk is set to be 1.0 initially.
  The size  of the clouds is set to be 5.0 $\times$ $10^{-3}$ in our units 
  unless specified.
  The radial and tangential restitution coefficient for cloud-cloud
  collisions are
  set to be 0.5 and
  0.0, respectively.

\subsubsection{Orbital configurations of galaxy mergers}
    In all of the merger  simulations, the orbit of the two disks is set to be
    initially in the $xy$ plane.
    The initial distance between
    the center of mass of the two disks,
    the pericenter
    distance,
    and the eccentricity of the merger orbit are 4.0, 1.0, and 1.0, respectively,
    for all models.
    The spin of each galaxy in a   merger
is specified by two angle $\theta_{i}$ and
    $\phi_{i}$, where suffix  $i$ is used to identify each galaxy.
    $\theta_{i}$ is the angle between the $z$ axis and the vector of
    the angular momentum of a disk.
    $\phi_{i}$ is the azimuthal angle measured from $x$ axis to
     the projection of the angular momentum vector of a disk on
    to $xy$ plane. 
$\theta_{1}$, $\theta_{2}$, $\phi_{1}$, and $\phi_{2}$ 
are set to be 30.0, 120.0, 90.0, and 0.0, respectively. 
The time when the progenitor disks merge completely and reach  the
dynamical equilibrium is less than 15.0 in our units for most of
models and does not depend so
strongly on the  history of star formation in the  present calculations.
In the present study, we do not describe the results
of models other than those given  above,
principally because 
diversity in initial orbital conditions of galaxy merging
only introduce a scatter in the present numerical results and 
does not drastically change the $essence$ of the derived results,  
as has been already pointed out  by Paper I.
Furthermore we do not here intend to investigate the importance of
multiple galaxy merging in the chemodynamical evolution of elliptical 
galaxies,
which has been already investigated in Paper I.

\subsubsection{Comparison with previous merger models}
 Initial conditions of galaxy mergers adopted in
 the present study differ from those adopted in previous studies
 (e.g., Mihos \& Hernquist 1996)
 mainly in the following two points.
 Firstly,
 as is described in \S 2.1.1, the present initial disk model does not
 include any remarkable bulge components, and accordingly corresponds to
 `pure'  late-type spiral. 
 The reasons for this  are firstly that
 by not including bulges, we can more clearly elucidate
 the dynamical effects of dissipative galaxy merging
 on chemical evolution of galaxies (that is,
 we can discriminate the effects of galaxy merging
 and those of galactic bulges),
 and secondly that extensive observational studies
 have not been yet accumulated which can give strong
 constraints on physical properties of galactic bulges
 in high redshift disks adopted in the present study.
 Although it is
 highly possible that  galactic bulges greatly affect the chemical evolution
 of galaxy mergers, we  investigate
 this issue in our future papers.
 Secondly, gas mass fraction adopted in the present study (1.0)
 is much larger that that of the observed typical values of
 the present disks ($0.1 \sim 0.2$, adopted in previous studies).
 This is firstly because we now consider chemodynamical evolution
 of gas-rich mergers at high redshift, and secondly because 
 we generally must begin with purely gaseous galactic disks for solving
 chemical evolution of galaxies.
 The main reason for our adoption of high redshift mergers
 is that the observed considerably tight color-magnitude relation of elliptical galaxies
 (Bower, Lucey, \& Ellis 1992; Ellis et al. 1997) and 
 the  relatively smaller  redshift evolution  of photometric properties
 of elliptical galaxies (Arag$\rm \acute{o}$n-Salamanca et al. 1993; Franx \& van Dokkum 1996)
 imply relatively  early formation of elliptical galaxies.
Actually, the gas mass fraction in precursor disks of a merger
is  different between galaxy mergers  and depends on the epoch of
the merging.
Although this difference  could yield a great variety of chemical
and dynamical structures in merger remnants, we do not intend to
consider this important difference for simplicity in the present  paper
and will address this in our future papers.

 Furthermore we should emphasize here the following three less realistic
 assumptions or oversimplifications on initial conditions of galaxy mergers
 adopted in the present study.
Firstly, in order to elucidate more clearly dynamical effects
on chemophotometric evolution of galaxy mergers,
we assume $fully$ $developed$ gas-rich exponential disks (`mature' disks)
in  investigating high redshift galaxy mergers.
It is, however, not observationally clarified  whether or not high redshift
disk galaxies are so fully developed (`mature') as is assumed in the present
study.
Accordingly, the adopted assumption of `mature' disks could
correspond to oversimplification.
Secondly, we neglect  the  importance 
of continuous gaseous infall onto galactic  disks from external
environments in the evolution of disk galaxies, which is expected
from observational studies on  
the G-dwarf problem in the Galaxy (e.g., van den Bergh 1962)
and star formation history
of galaxies (e.g., Larson 1972; 
Larson, Tinsley, \& Caldwell 1980; Kennicutt 1983; Sandage 1986).
Owing to this neglect of gaseous infall for merger progenitor disks,
the present chemodynamical model suffers from overproduction
of metal-poor stars in merger remnants, as is pointed out by  
other chemical evolution models (e.g.,  Tantalo et al. 1996).
Thirdly, we assume that two disks merge with each other to form 
an elliptical when the disks still have a larger gas mass
fraction ($\ge 0.2$).
Owing to this assumption, the dynamical effects of galaxy merging
on ISM and thus on chemical evolution of ISM is more drastic
in the present chemodynamical model.
Actually, the epoch of major galaxy merging is probably widely spread,
depending on orbit configurations of galaxy merging (impact
parameters) and galaxy environments.
Accordingly, it is equally possible that in gas-poor galaxy mergers
(late mergers),
galaxy merging does not induce the secondary starburst strongly
enough to change substantially stellar populations of galaxy mergers.
Because of these three  apparently less realistic assumptions,
we should interpret  more carefully the present
numerical results and consider the applicability of
the results to real elliptical galaxies.
These assumptions indeed  may weakens the validity
of the present chemodynamical model, 
however, we believe  that even the present model enables us to
grasp $some$ essential ingredients of dynamical effects on chemophotometric 
evolution
of galaxy mergers.
Our future studies, which do not adopt  the above
three  less realistic assumptions
and instead include more realistic initial conditions
of galaxy mergers within a plausible cosmological model,
will help us to understand more deeply and thoroughly 
galactic chemophotometric evolution affected largely by
dynamical evolution of galaxy mergers.

\subsection{Star formation model}
\subsubsection{Gas consumption rate}

    Gas consumption by star formation in galaxy mergers
     is modeled by converting  the collisional
    gas particles 
    into  collisionless new stellar ones  according to the algorithm
    of star formation   described below.
    We adopt the Schmidt law (Schmidt 1959)
    with exponent $\gamma$ = 2.0 (1.0  $ < $  $\gamma$
      $ < $ 2.0, \cite{ken89}) as the controlling
    parameter of the rate of star formation.
    The amount of gas 
    consumed by star formation for each gas particle
    in $each$ $time$ $step$, 
    $\dot{M_{\rm g}}$, 
is given as:
    \begin{equation}
      \dot{M_{\rm g}} \propto  C_{\rm SF} \times 
 {(\rho_{\rm g}/{\rho_{0}})}^{\gamma - 1.0}
    \end{equation}
    where $\rho_{\rm g}$ and $\rho_{0}$
    are the gas density around each gas particle and
    the mean gas density at 0.48 $R_{\rm d}$  of 
    an initial disk, respectively.
    This star formation model is similar to that of Mihos, Richstone, \& Bothun  (1992).
    In order to avoid a large number of new stellar particles with
different mass, we convert one gas particle into one stellar one
according to the following procedure.
First we give each gas particle the probability, $P_{\rm sf}$,
 that the gas particle
is converted into stellar one, by setting the $P_{\rm sf}$ to be proportional
to the  $\dot{M_{\rm g}}$  in equation (1) estimated for the gas particle. 
Then we draw the random number to determine whether or not the gas particle
is totally converted  into one new star. 
This method of star formation enables us to control the rapidity of star formation
without increase of particle number in each simulation thus to maintain
the numerical accuracy in  each simulation. 
The  $C_{\rm SF}$ is meant to be  proportional to
the ratio of dynamical time-scale of the system to the time-scale of
gas consumption  by star formation.
We accordingly emphasize that even if the coefficient of the Schmidt law
is constant for galaxies, values of $C_{\rm SF}$ can be  different
between galaxies owing to the difference  in dynamical time-scale between 
galaxies.
The $C_{\rm SF}$ is considered to depend exclusively on 
galactic disk mass ($M_{\rm d}$), as is described later.
The equation (1) furthermore states that
a larger number of stellar particles are created at the
regions  where the local gas density becomes larger owing to the onset
of local Jeans instability.
The positions and velocity of the new stellar particles are set to 
be the same as those of original gas particles.

\subsubsection{Supernovae feedback effects}

Thermal and dynamical (kinematic) heating of ISM  driven by
supernovae are found to affect greatly  the formation and
evolution of galaxies
(e.g., Katz 1992; Navarro \& White 1993). 
We here consider  the dynamical  feedback effects of Type Ia 
(SNIa) and Type II supernovae (SNII) and neglect the feedback effects
of thermal heating on galaxy evolution.
The reason for this neglect
is firstly that our gaseous model does not solve thermal evolution
of ISM, and secondly that  such thermal effects is found to be less
important than dynamical feedback effects owing to efficient cooling
of ISM (e.g., Katz 1992; Navarro \& White 1993).
Accordingly some fraction of  total energy ejected from SNIa and SNII is  assumed to  
give velocity perturbation of  ISM in the present study.
The details of the supernovae feedback effects are given as follows.
Total energy produced by SNIa and SNII at the time $t$ for
each $i$th newly born stellar particle ($\Delta E_{\rm sn, \it i} (t)$)
is given as,
\begin{equation}
\Delta E_{\rm sn, \it i} (t) = m_{\rm s, \it i}S_{i}(t-t_{i}), 
\end{equation}
where  $m_{\rm s, \it i}$ is the mass of the $i$th stellar particle,
$S_{i}(t-t_{i})$ is the total energy of SNI and SNIIa per unit mass
at the time $t$,
and $t_{i}$ represents the time when the $i$th stellar particle is
born from a gas particle.
The above formalism is more elaborated than previous studies in the
following two points.
Firstly, we consider  time-delay between the epoch of star formation
and the epoch of supernovae explosions for stars with variously different masses.
In general, life-time of stars depends on stellar mass in such a way that
more massive stars die earlier than less massive ones.
Consequently, the time-delay is different between stars with different masses
and total number of SNIa and SNII  is also time-dependent.
Since each $i$th stellar particle ($\sim 10^{6} M_{\odot}$ in the present study)
is assumed to consist of stars with different masses,
the $S_{i}(t-t_{i})$ is  time-dependent.
Secondly, we consider that both the energy ejected from SNIa and that
from SNII can be returned to ISM of galaxy mergers but neglect
thermal energy ejected from long-lived evolved stars.
The total energy per a supernovae is assumed to be $10^{51}$erg both
for SNIa and SNII.
The typical epoch of supernovae explosions are different between
these two types of supernovae ($\sim 10^{9}$ yr for SNIa
and $\sim 10^{7}$ yr for SNII),
and  this difference is reflected on  the time-dependence
of the $S_{i}(t-t_{i})$.

Some fraction of the $\Delta E_{\rm sn, \it i} (t)$ are assumed to be
returned to gas particles (ISM)  and then to change
kinetic energy of gas particles surrounding  $i$th  stellar particle.
Each $j$th gas particle around $i$th  stellar particle can
receive a velocity perturbation ($\Delta v_{\rm sn, \it j}$) directed radially away from 
the $i$th  stellar particle. The $\Delta v_{\rm sn, \it j}$ at the time $t$ satisfies
the following relation:
\begin{equation}
f_{\rm sn}\Delta E_{\rm sn, \it i} (t) = 
 1/2\sum_{j}^{N_{\rm nei, \it i}} m_{\rm g, \it j} {(\Delta v_{\rm sn, \it j})}^{2} ,
\end{equation}
where $f_{\rm sn}$ represents 
the fraction of supernovae energy returned to
ISM, $N_{\rm nei, \it i}$ is total number of gas particles that
are located within 0.05 (in our units) around $i$th stellar particle,
and $m_{\rm g, \it j}$ is a mass of each $j$th gas particle.
This formulation  is similar to that adopted in Navarro \& White (1993).
It depends largely on 
the parameter $f_{\rm sn}$ how greatly SNIa and SNII can provide
dynamical (kinetic) feedback effects on ISM of galaxy mergers.
It is, however,  not observationally clarified which values of $f_{\rm sn}$
is the most plausible and realistic for the evolution of ISM
affected by SNIa and II feedback.
Thus we investigate the importance of the parameter $f_{\rm sn}$
by varying values of $f_{\rm sn}$ ($0.0 \le f_{\rm sn} \le 0.5$
in the present study).

   All the calculations related to 
the above dynamical evolution  including the dissipative
dynamics, star formation,  SNIa and SNII feedback effects,
and gravitational interaction between collisionless
and collisional component 
 have been carried out on the GRAPE board
   (\cite{sug90})
   at Astronomical Institute of Tohoku University.
  The  number of particles
  for a galaxy merger model is 
  10000 for the halo component and 20000 for the gaseous component.
   The parameter of gravitational softening is set to be fixed at 0.04  
   in all the simulations. The time integration of the equation of motion
   is performed by using 2-order
   leap-flog method. Energy and angular momentum  are conserved
within 1 percent accuracy in a test collisionless merger simulation.
Most of the  calculations are set to be stopped at T = 20.0 in our units
unless specified.

\subsection{Chemophotometric model}

Photometric and spectroscopic properties such as global colors and metallicity
gradient in elliptical galaxies
depend critically on how the metallicity and age of  stellar components
are distributed in  galaxies.
Accordingly, we must analyze the distribution of stellar age and 
that of stellar metallicity in merger remnants for each model in
order to grasp some fundamental  characteristics of 
chemical, photometric, and spectroscopic properties
in merger   remnants. 
We calculate
age and metallicity distribution,
based on the age ($t_{i}$) and metallicity ($Z_{i}$) assigned to  
each $i$th stellar particle, as described in detail later.
The outline for this calculation is as follows. 
First we derive the  distribution of  stellar age and that of stellar 
metallicity 
by assigning  age and metallicity to stellar particles according as the
law of chemical enrichment applied to this study.
Next, by using the stellar  population synthesis method,
we calculate  the photometric and spectroscopic properties 
based on the derived distribution
of age and metallicity in merger remnants.

\subsubsection{Chemical enrichment}

 Chemical enrichment through star formation during galaxy merging
is considered  to proceed  locally and inhomogeneously.
In the present study, we investigate time-evolution
of  five species of chemical components,
H, He, O, Mg, and Fe as well as the conventionally used mean
metallicity $Z$. 
The mean metallicity of $Z$
for each $i$th stellar particle is represented by $Z_{i}$. 
Total mass of each $j$th  ($j$=1,2,3,4,and 5) chemical component (H, He, O, Mg, and Fe) 
ejected from  each $i$th stellar particle at the time $t$ 
is given as, 
\begin{equation}
\Delta z_{i,j} (t) = m_{\rm s, \it i}Y_{i,j}(t-t_{i}), 
\end{equation}
where  $m_{\rm s, \it i}$ is the mass of the $i$th stellar particle,
$Y_{i,j}(t-t_{i})$ is the mass of each $j$th chemical component  ejected
from stars per unit mass
at the time $t$,
and $t_{i}$ represents the time when the $i$th stellar particle is
born from a gas particle.
The $\Delta z_{i,j} (t)$ is given to
neighbor gas particles locating within
$R_{\rm chem}$ from the position of the $i$th  stellar particle.
This  $R_{\rm chem}$ is referred to as 
chemical mixing length in the present paper,
and represents the region within which the neighbor
gas particles are polluted by metals ejected from  stellar particles.
The value of  $R_{\rm chem}$  relative to the typical size of
a galaxy could be different between galaxies, accordingly the
value of  $R_{\rm chem}$  is considered to be a free parameter in
the present study.
The value of  $R_{\rm chem}$ examined the most extensively 
in the present study is 
0.1, which corresponds to the half of the scale-length of initial disks.
Initial gaseous metallicity  ($Z$)
is set to be 0.1 solar, which is exactly the same as that
of infall gas adopted in the chemical evolution of disk galaxies
with gaseous infall (e.g., Matteucci \& Fran\c ois 1989).
We adopt this initial enrichment of ISM in disks in order to avoid
overproduction of less metal-enriched stellar components that are 
not allowed for plausible chemical evolution of disk galaxies (e.g., van den Bergh 1962). 
Initial value of the abundance ratio [Mg/Fe] for ISM in disks
is set to be 0.8, which is similar to a typical value of metal-poor
halo ISM that is metal-enriched mostly by SNII and a candidate for infall gas
of disk galaxies (e.g., Matteucci \& Fran\c ois 1989).

Since we now consider the time-delay between the epoch of star formation
and that of supernovae explosions, the mass
of each $j$th chemical component ejected from each
$i$th stellar particle,
which are basically  determined  by $Y_{i,j}(t-t_{i})$,
is also strongly time-dependent.
We  estimate the mass-dependent life-time 
of stars that becomes SNIa or SNII 
by using mass-age relation given by Bressan et al. (1993).
The fraction of close binary stars 
of SNIa relative to SNII 
(represented by  $A$ parameter
in Matteucci \& Tornamb\`{e} 1987)
is assumed to be 0.1.
The $Y_{i,j}(t-t_{i})$ furthermore depends on stellar yields, IMF profiles,
upper cutoff mass $M_{\rm up}$, and lower one $M_{\rm low}$.
In the present study, 
we adopt the Salpeter  IMF,
$\phi(m) \propto m^{-1.35}$,
with $M_{\rm up}=120 M_{\odot}$ and
$M_{\rm low}=0.6 M_{\odot}$.
The reason for this larger $M_{\rm low}$ (=$0.6 M_{\odot}$)
is essentially because we do not have stellar yield tables
for stars with  masses less than 0.6.
To calculate the ejected mass of gas and metals in $Y_{i,j}(t-t_{i})$,
we use stellar yields derived by  Woosley \& Weaver (1995) for SNII,
Nomoto, Thielemann, \& Yokoi (1984) for  SNIa,
and Bressan et al. (1993)
and  Magris \& Bruzual (1993) for low and intermediate mass stars.
More details of the time-dependence of $Y_{i,j}(t-t_{i})$ for a given
IMF, $M_{\rm up}$, and $M_{\rm low}$ are  given in Appendix A.

\subsubsection{Population synthesis}

It is assumed in the present study
that the spectral energy distribution (SED) of a model galaxy is 
a sum of the SED of  stellar particles. 
The SED of each  $i$th stellar particle is assumed to be  
a simple stellar population (SSP) that is  
a coeval and chemically homogeneous assembly of stars. 
Thus the monochromatic flux of a galaxy with  age $t$,
$F_{\lambda}(t)$,  is described as; 
\begin{equation}
F_{\lambda}(t) = \sum_{\rm star} F_{{\rm SSP},\lambda}(Z_{i},
{\tau}_{i}) \times m_{\rm s, \it i} \; ,
\end{equation}
where $F_{{\rm SSP},\lambda}(Z_{i},{\tau}_{i})$ and $m_{\rm s, \it i}$ 
 are  a monochromatic flux of SSP 
of age ${\tau}_{i}$ and metallicity $Z_{i}$, where suffix $i$ identifies 
each $i$th stellar particle,  and 
mass of each stellar  particle,  respectively.
The age of SSP, ${\tau}_{i}$, is defined as ${\tau}_{i} = t - t_{i}$, 
where $t_{i}$ is the time when a gas particle is converted into a stellar one.
The metallicity of SSP is exactly the same
as that  of the stellar particle, $Z_{i}$, and the summation ($\sum$) in
equation (5) is done  for all 
stellar particles in a model galaxy.
A stellar particle is assumed to be composed of stars whose
age and metallicity are exactly the same as those of the stellar particle
and 
the total mass of the stars is set to be the same as that of
the  stellar particle.
Thus the monochromatic flux of SSP at a given wavelength is defined as
\begin{equation}
F_{{\rm SSP}, \lambda}(Z_{i},{\tau}_{i}) = \int_{M_{\rm low}}^{M_{\rm up}} 
\phi (M) f_{\lambda}(M, {\tau}_{i}, Z_{i}) dM \; ,
\end{equation}
where $M$ is mass of a star, $f_{\lambda}(M, {\tau}_{i}, Z_{i})$
 is a monochromatic flux 
of a star with mass $M$, metallicity $Z_{i}$ and age ${\tau}_{i}$,
and 
$\phi (M)$ is a initial mass function (IMF) of stars.
In this paper, we use the $F_{{\rm SSP}, \lambda}(Z_{i}, {\tau}_{i})$ 
of  GISSEL96 which is a latest version of Bruzual \& Charlot (1993).

\placetable{tbl-1}

\subsection{Main points of analysis}
 We mainly investigate how initial galactic disk mass ($M_{\rm d}$)
 determines chemical,  photometric, and spectroscopic  evolution of galaxy mergers
 by varying values of the parameter $M_{\rm d}$. 
 In the present study, 
 we assume Freeman's law (Freeman 1970),
 $M_{\rm d} \propto {R_{\rm d}}^{2}$, and constant
 initial ratio of dark matter to disk mass ($M_{\rm d}/M_{\rm t} \sim$ constant).
 Hence if we give $M_{\rm d}$, then $R_{\rm d}$, $v$, and $t_{\rm dyn}$ are
 automatically determined 
 ($R_{\rm d}$, $v$, and $t_{\rm dyn}$ are
 described as a function of $M_{\rm d}$.).
 We here adopt $M_{\rm d}$ = 6.0 $\times$ $10^{10}$ $  M_{\odot}$,
 $R_{\rm d}$ = 17.5 kpc,
 $v$ = 1.21 $\times$
 $10^{2}$ km/s,  and  $t_{\rm dyn}$ = 1.41 $\times$ $10^{8}$ yr
 as  reference  values,
 which are exactly the same as those in Paper I.
 The $M_{\rm d}$ dependence of $R_{\rm d}$, $v$, and $t_{\rm dyn}$ are 
 accordingly described as:
 $R_{\rm d} = 17.5 {(M_{\rm d}/6.0 \times 10^{10}   M_{\odot})}^{1/2} $kpc,
 $v =  1.21 \times 10^{2} {(M_{\rm d}/6.0 \times 10^{10})}^{1/4}$ km/s,
 $t_{\rm dyn} = 1.41 \times 10^{8} {(M_{\rm d}/6.0 \times 10^{10})}^{1/4}$ yr.
 We here emphasize that the adopted  scaling is only fiducial,
 thus absolute magnitude of the  present results based on this scaling can be changed if
 other types of galactic scaling are adopted.
 $M_{\rm d}$ is considered to be the most important parameter in the present study,
 principally because $M_{\rm d}$ can control the following two physical quantities
 that greatly affect galactic evolution.
 The first is 
 the rapidity of star formation,
 which means the ratio of dynamical time-scale to star formation one
represented by $C_{\rm SF}$ in \S 2.2.1.
 The second is that  the effectiveness of supernovae feedback,
 which means the ratio of total energy expected to be ejected from
 supernovae (in the course of galaxy evolution) 
 to total potential energy in a galaxy (hereafter
 represented by $f_{\rm eff}$).
 The mass-dependent properties of galaxy mergers,
 which are described later (\S 3),
 are due essentially to mass-dependence of these
 two quantities ($C_{\rm SF}$ and $f_{\rm eff}$)
 in the present study.
 The expected $M_{\rm d}$  dependence of $C_{\rm SF}$ and $f_{\rm eff}$
 are $C_{\rm SF} \propto (M_{\rm d})^{0.25} $ and
 $f_{\rm eff} \propto (M_{\rm d})^{-0.5} $.
 Accordingly, if we set both the  value of $C_{\rm SF}$ and
 that of $f_{\rm eff}$ in the model with  $M_{\rm d}=10^{10} M_{\odot}$
 to be  1.0,
 the ranges of the two quantities 
 are $1.0 \le C_{\rm SF} \le 3.16$ and $1.0 \le  {f_{\rm eff}}^{-1} \le 4.0$
 for galaxy mergers with $10^{10} M_{\odot} \leq M_{\rm d} \leq 10^{12} M_{\odot}$.
 More details of these dependence 
are given in Appendix B.

  We particularly investigate $M_{\rm d}$ dependence
  of $mean$ and $radail$ properties 
  of stellar metallicity (Fe, Mg, and Z),
  the abundance ratio of [Mg/Fe],
  global colors ($U-R$, $B-R$, and $V-K$),
  and line indices ($\rm H_{\beta}$,  $\rm Mg_{2}$, and 
$\langle Fe \rangle$),
  mass-to-light-ratio ($M/L_{B}$ and $M/L_{K}$).
  Here $\langle Fe \rangle$ represents the mean value of Fe5270 and Fe5335 (Faber et al. 1995).
  The values of $M_{\rm d}$
  range from $10^{10}  M_{\odot}$ to $10^{12}  M_{\odot}$
  in the present merger models.
  We furthermore investigate fundamental roles of  
  chemical mixing length
  ($0.1 \leq R_{\rm chem} \leq 0.4$) and fraction of supernovae energy that is returned to
  ISM ($0.0 \leq f_{\rm sn} \leq 0.5$).
  In total,  13  model  are investigated, including models with different masses (Model 1 $\sim$ 6),  
  those with no supernova feedback (Model 7, 8, and 9),
  those with larger chemical mixing length (Model 10 and 11),
  and those with no gaseous dissipation (Model 12 and 13).
  The values  of parameters and brief summary of the results are given in Table 1.
  First, second,  third,  and fourth columns
  in the Table 1 denote the model number,
  the value of $M_{\rm d}$, that of $R_{\rm chem}$, and that of $f_{\rm sn}$ 
  respectively, for each model.
  Fifth, sixth, seventh, eighth, ninth, and tenth  columns give mean stellar metallicity
  ($\langle Z_{\ast} \rangle$), the abundance ratio of [Mg/Fe],
  global $V-K$ color (in units of mag), $\rm Mg_{2}$ (mag), 
  $\rm H_{\beta}$ (\AA), and $\langle {\rm Fe} \rangle$ 
  line indices (\AA) in the central region
  of merger remnants (within  $0.1R_{\rm e}$).
  Mass-to-light ratio, which means the ratio of total $stellar$ mass
  to stellar luminosity, is given in eleventh column 
  for  $B$-band ($M/L_{\rm B}$) and 
  in twelfth one for $K$-band ($M/L_{\rm K}$). 
  Thirteen column denotes
  the value of metallicity gradient 
($\Delta \log \langle Z_{\ast} \rangle/\Delta \log R$).
  Comments for each model are given in the fourteenth column.
  As is described above, 
  our main points of the present study are not the origins of structural and kinematical
  properties characteristics of ellipticals.
  Therefore we do not here intend to discuss these properties.
  More details on fundamental structural and kinematical properties of merger remnants are given
  in Bekki \& Shioya (1997) and our Paper I.
  Lastly, it  should be stressed  that although
  the $relative$ $importance$ of each parameter in determining galactic properties
  derived in the present study 
  (e.g.,  $M_{\rm d}$ dependence of chemophotometric properties of merger remnants)
  is considerably meaningful, 
  the $absolute$ $magnitude$ of the derived galactic properties
  should be more carefully interpreted:
  This is primarily because the present chemodynamical model
  is still rather idealized in  terms of dissipative dynamics, star formation
  model, time-independent IMF, chemical enrichment processes of galaxy mergers,
  and galaxy scaling relations.

\placefigure{fig-1}
\placefigure{fig-2}
\placefigure{fig-3}

\section{Results}

Firstly, we describe how initial disk mass of galaxy mergers
determines fundamental chemical, photometric, and spectroscopic
properties of merger remnants in \S 3.1.
Secondly, we emphasize the importance of supernova feedback
effects in the formation of spectrophotometric 
relationships observed in  elliptical galaxies in \S 3.2.
Thirdly, we provide fundamental roles  of parameters other than
galactic mass in determining spectrophotometric
properties of merger remnants in \S 3.3.

\subsection{Dependence on initial disk mass of galaxy mergers}

\subsubsection{Metallicity distribution}

Figure 1 shows time-evolution of gas mass fraction ($f_{\rm g}$)
and that of mean stellar metallicity ($\langle Z_{\ast} \rangle$) 
in galaxy mergers for Model 1 with $M_{\rm d}=10^{10} M_{\odot}$
and  Model 2 with $M_{\rm d}=10^{12} M_{\odot}$.
A larger amount of ISM is finally converted into stellar components
in a more massive galaxy merger, firstly because
a smaller amount of ISM is tidally stripped away from the merger
owing to more rapid star formation, and secondly because
the supernovae feedback does not so strongly suppress the formation
of higher gaseous density region that is a potential site of efficient
star formation in the merger.
Consequently, a more massive galaxy merger has larger mean stellar metallicity,
as is shown in Figure 1.
This dependence of mean stellar metallicity on galactic mass is consistent
with that derived in our Paper I.
Figure 2  describes final metallicity distributions of stellar components ($Z_{\ast}$) 
in merger remnants for Model 1 with $M_{\rm d}=10^{10} M_{\odot}$
and Model 2 with $M_{\rm d}=10^{12} M_{\odot}$.
The peak of the metallicity distribution is located in the righter side for
a more massive merger remnant, which means a larger amount of stellar components
is substantially metal-enriched in the remnant.
Figure 3 gives time evolution of stellar metallicity, 
$\langle Z_{\ast, \rm Fe} \rangle$,
$\langle Z_{\ast, \rm Mg} \rangle$ and 
that of stellar abundance ratio of [Mg/Fe].  
As is shown in Figure 3,
both $\langle Z_{\ast, \rm Fe} \rangle$ and
$\langle Z_{\ast, \rm Mg} \rangle$ are finally discernibly larger  for a more  massive
merger.
This is essentially because for the massive galaxy merger,
owing to less effective supernova feedback,
a larger amount of ISM can  be converted into stellar components 
during dissipative galaxy merging and thus more metal-enriched with Mg and Fe.
Furthermore  the stellar abundance ratio of [Mg/Fe]
is slightly  smaller  for the more massive merger remnant.
It should be emphasized here that [Mg/Fe] is larger than the solar ratio ([Mg/Fe]=0.0)
for the two  merger models.
This result reflects the fact that in the present merger model,
two gas-rich disks are assumed to  merge with each other at high redshift universe,
where merger progenitor disks have a larger amount of less-metal enriched ISM.

\placefigure{fig-4}
\placefigure{fig-5}
\placefigure{fig-6}
\placefigure{fig-7}

\subsubsection{Radial gradients}
 The present merger model demonstrates  that
 chemical evolution of ISM in galaxy merges proceeds in  an very inhomogeneous
 way: In the central part of mergers, metal-enrichment is more efficient
 owing to radial inflow of metal-enriched ISM and the subsequent star formation
 triggered by dissipative galaxy merging, whereas in the outer part,
 metal-enrichment is less efficient owing to  a larger amount of metal-enriched
 ISM tidally
 stripped away from mergers.
 As a natural result of this inhomogeneous chemical mixing in dissipative galaxy merging, 
 both stellar and gaseous
 components have a number of interesting 
 chemical (and spectrophotometric) properties in merger remnants.  
 Among these, we  here  restrict ourselves to radial properties
 of stellar populations, and accordingly other important roles of inhomogeneous
 chemical mixing in determining fundamental properties of elliptical galaxies
 are described in \S 4.2.
 Figure 4 describes radial distribution of stellar metallicity
 ($\langle Z_{\ast} \rangle$, $\langle Z_{\ast, \rm Fe} \rangle$, 
and $\langle Z_{\ast, \rm Mg} \rangle$)
 and that of the abundance ratio of [Mg/Fe] in merger remnants
for Model 1 with $M_{\rm d}=10^{10} M_{\odot}$
and Model 2 with $M_{\rm d}=10^{12} M_{\odot}$.
Irrespectively of galactic mass,
merger remnants show $negative$ metallicity gradients fitted well by
power-low profile (especially for the region $0.5 \leq R/R_{\rm e} \leq 5.0$).
The metallicity gradients 
($\Delta \log \langle Z_{\ast} \rangle / \Delta \log R$)
are steeper  in the more massive merger remnant for  the region 
$0.1 \leq R/R_{\rm e} \leq 1.0$: 
$\Delta \log \langle Z_{\ast} \rangle / \Delta \log R$ is  $-0.03$ 
for Model 1 and $-0.13$ for Model 2.
This dependence is true for radial gradients of 
$\langle Z_{\ast, \rm Fe} \rangle$ and $\langle Z_{\ast, \rm Mg} \rangle$.
The physical reason for this mass-dependence of metallicity gradients 
is principally  that for a more massive galaxy merger,
supernova feedback does not so strongly suppress the inward transfer
of more metal-enriched ISM during dissipative galaxy merging
and accordingly 
more metal-enriched ISM
can be transferred to the central region of the merger and converted
into new stellar components there.
The derived dependence of stellar metallicity gradients on galactic mass
is considerably different from that described in our Paper I, in which
supernova feedback effects are not included at all.
This difference between the present study and our Paper I strongly
suggests that supernova feedback effects are critically important
for the origin of stellar metallicity gradients of elliptical galaxies.
The radial gradient of [Mg/Fe], on the other hand,
shows flat profile in the inner regions of merger remnants for the two models.
Figure 5 shows the radial gradient of mean epoch of star formation
($\langle T_{\rm SF} \rangle$) in merger remnants 
for Model 1 with $M_{\rm d}=10^{10} M_{\odot}$
and Model 2 with $M_{\rm d}=10^{12} M_{\odot}$.
Here,  a larger value of $\langle T_{\rm SF} \rangle$ at a given point  means
that stellar populations at the point are younger in average.
Considering the results in Figure 4,
Figure 5 clearly demonstrates that irrespectively of galactic mass,
stellar populations of merger remnants are both younger and metal-enriched
in the central part of merger remnants.
This result reflects the fact that in the central part of galaxy mergers,
more metal-enriched ISM that is transferred to galactic nuclei
owing to efficient gas fueling  in the later stage
of galaxy merging 
can be preferentially and efficiently converted into stellar
components within the deeper potential well. 
These  younger and more metal-enriched  stellar populations in the
central part of merger remnants 
are  consistent reasonably well with recent observational results of elliptical
galaxies by
Faber et al (1995).

Figure 6 describes radial gradients of integrated colors
($U-R$, $B-R$, and $V-K$) in merger remnants for Model 1 and 2.
The color gradients 
($\Delta U-R / \Delta \log R$, $\Delta B-R / \Delta \log R$, 
$\Delta V-K / \Delta \log R$)
are more likely to be steeper  in the more massive merger remnant for  the region 
$0.1 \leq R/R_{\rm e} \leq 1.0$: 
For example, $\Delta U-R / \Delta \log R$ is  $-0.07$ for Model 1
and $-0.13$ for Model 2.
This mass-dependence results from the already derived
dependence that more massive merger remnants have larger metallicity gradients.
The derived color gradients are appreciably shallower than mean values of those observed
by  Peletier, Valentijn, \& Jameson (1990) and Peletier et al. (1990). 
We here emphasize that if we measure the color gradients for the region
$1.0 \leq R/R_{\rm e} \leq 10.0$, even color gradients of less massive
merger remnants can become larger owing to considerably less metal-enriched 
stellar populations in the outer part of the merger remnants.
Figure 7 gives radial gradients of three line indices 
($\rm H_{\beta}$, $\rm Mg_{2}$, and $\rm \langle Fe \rangle$) in merger remnants for
Model 1 and 2.
The  radial gradients of $\Delta \rm H_{\beta} / \Delta \log \it R$, 
$\Delta \rm Mg_{2} / \Delta \log \it R$, 
and $\Delta \rm \langle Fe\rangle / \Delta \log \it R$ 
for the region $0.1 \leq R/R_{\rm e} \leq 1.0$
are 0.036, $-0.013$, and $-0.121$, respectively, for Model 1,
and 0.051, $-0.025$, and $-0.250$, respectively, for Model 2.
The physical reason for the $positive$ gradient in $\rm H_{\beta}$ line index
is that stellar populations in the outer part of merger remnants
are considerably less metal-enriched.
The derived  gradients of line indices are appreciably shallower than mean values of those observed
by  Davies, Sadler, \& Peletier (1990). 
Moreover the radial gradients of $\rm Mg_{2}$ and $\rm \langle Fe \rangle$
are steeper for a more massive merger remnant,
which means that a larger amount of metals ejected from SNIa and SNII 
are transferred to the central part of the remnant owing
to more efficient fueling of metal-enriched ISM
for the more massive galaxy merger.

\placefigure{fig-8}
\placefigure{fig-9}
\placefigure{fig-10}

\subsection{Supernova feedback effects and 
spectrophotometric properties of merger remnants}

Here in \S 3.2, we present  dependence of
spectrophotometric properties of merger remnants
on the parameter $f_{\rm sn}$.
We compare the results of models with $f_{\rm sn}=0.0$ (no supernova
feedback models, Model 7, 8, and 9) with those of models with $f_{\rm sn}$ =0.1 and 0.5
(Model 1 $\sim$ 6)
and thereby observe how the supernova feedback effects
control fundamental characteristics of spectrophotometric properties 
in merger remnants.
In particular, we  describe the importance of the supernova feedback effects
in the formation of the following three  observational spectrophotometric
relationships of elliptical galaxies: (1) The CM relation, 
(2) The $\rm Mg_{2}$-$\sigma$  relation,
and (3) The luminosity-dependence of mass-to-light ratio.
We here focus on rather $qualitative$ behaviors of the derived results,
and accordingly  the $qualitative$ comparison of the present results
with observational ones is given in \S 4.1. 

\subsubsection{The CM relation}
Figure 8 shows the dependence of $U-R$, $B-R$, and $V-K$ colors
on $B$ band luminosity ($L_{\rm B}$) in merger remnants
for models with different initial disk mass ($M_{\rm d}=10^{10},
10^{11}, 10^{12} M_{\odot}$) and  strength of supernovae
feedback ($f_{\rm sn}=0.0,0.1,0.5$).
We can clearly observe the color-magnitude relation of merger remnants 
for  each color ($U-R$, $B-R$, and $V-K$) in Figure 8:
More luminous merger remnants (ellipticals) have redder colors.
The primary reason for the derived color-magnitude relation
of merger remnants is that more massive (luminous) galaxy mergers
can have larger mean stellar metallicity owing to more efficient star
formation and chemical enrichment:
The CM relation derived in the present study
is due to mass-metallicity relation of merger remnants.
This result implies 
that the existence of the observed  
CM relation is not necessarily inconsistent with the merger scenario
of elliptical galaxy formation,
furthermore that the origin of the observed CM relation
can be closely associated with dissipative galaxy merging.
As is observed in Figure 8,
the difference of colors between merger remnants with different 
luminosity can be more discernable  in the models with $f_{\rm sn}=0.5$
than in the models with $f_{\rm sn}=0.0$ (no feedback).
This result suggests that in addition to the rapidity
of star formation (the importance of which is described in the Paper I),
supernova feedback effects play a vital role
in reproducing the CM relation of elliptical galaxies.
Thus more luminous ellipticals formed by galaxy mergers between more massive spirals
show redder colors, principally because more massive mergers
suffer less severely from supernovae feedback effects thus have more metal-enriched
stellar populations.

\subsubsection{The $\rm Mg_{2}$-$\sigma$  relation}

 Figure 9 gives the dependence of central $\rm Mg_{2}$ index (within  $0.1 R_{\rm e}$)
on the central velocity dispersion 
in merger remnants
for models with different initial disk mass ($M_{\rm d}=10^{10},
10^{11}, 10^{12} M_{\odot}$) and  strength of supernovae
feedback ($f_{\rm sn}=0.0,0.1,0.5$).
As is observed in Figure 9,
although there is no clear trend in $\rm Mg_{2}$-$\sigma$  relation
for models with $f_{\rm sn}=0.0$ (no supernova feedback),
$\rm Mg_{2}$-$\sigma$  relation which is qualitatively consistent
with that observed in elliptical galaxies  is  more discernable for models
with $f_{\rm sn}=0.1$ and 0.5.
This result clearly demonstrates that the origin of the observed
$\rm Mg_{2}$-$\sigma$ relation of elliptical galaxies can
be due partly to  physical processes associated with SNIa and SN II events.
The reason for the clearer $\rm Mg_{2}$-$\sigma$  relation in models
with $f_{\rm sn}=0.1$ and 0.5 is essentially that
for more massive galaxy mergers,
a larger amount of more metal-enriched ISM can be transferred to the central
part of the  mergers with deeper potential wells.
Moreover the $\rm Mg_{2}$-$\sigma$  relation derived in models with $f_{\rm sn}=0.5$
is more stepper and thus similar to the observational one
than that derived in models with $f_{\rm sn}=0.1$.
This result implies that stronger supernovae feedback effects can be required
for more successful reproduction of the observed $\rm Mg_{2}$-$\sigma$  relation.
Thus, the present chemodynamical model 
suggests that  more massive ellipticals with larger central velocity
dispersion formed by galaxy mergers between more massive gas-rich spirals 
are more likely to show larger central  $\rm Mg_{2}$ index.

\subsubsection{The mass-to-light-ratio}

 Figure 10 shows dependence of `mass-to-light-ratio' (in $B$ and $K$ bands)
 on luminosity in merger remnants
for models with different initial disk mass ($M_{\rm d}=10^{10},
10^{11}, 10^{12} M_{\odot}$) and  strength of supernovae
feedback ($f_{\rm sn}=0.0,0.1,0.5$).
Here the `mass-to-light-ratio' means the ratio of total stellar mass
to total stellar luminosity, thus does not mean the ratio of
total mass (stellar mass and dark matter mass) to total stellar luminosity.
As is shown in Figure 10, mass-to-light ratio in $B$ band ($M/L_{B}$) is larger
for more luminous merger remnants whereas that in $K$ band ($M/L_{K}$) is
smaller for more luminous merger remnants.
The essential reason for these dependences is that
the more luminous merger remnants have larger mean stellar 
metallicity owing to more efficient  chemical evolution
of galaxy mergers.
Furthermore, the luminosity dependence of mass-to-light-ratio
can be more discernibly observed in the models with $f_{\rm sn}=0.1$ and 0.5
than in those with $f_{\rm sn}=0.0$ (no feedback effects).
This is primarily because the difference in mean stellar metallicity 
between galaxy mergers with different masses is larger in models with $f_{\rm sn}=0.1$ and 0.5
than in those with $f_{\rm sn}=0.0$.
Assuming that the observed mass-to-light ratio inferred from the  FP
is the ratio of stellar mass  to stellar luminosity (not
the ratio of total mass to stellar luminosity), 
the derived dependence of $M/L_{B}$ on $L_{B}$ is only qualitatively
consistent with that inferred from the FP of elliptical galaxies
(e.g., Djorgovski \& Davis 1987; Dressler et al. 1987; Bender et al. 1992;
Pahre, de Carvalho, \& Djorgovski 1998).
This result implies that the effect of stellar population alone (that is, the mass-metallicity
relation)  can not explain
the origin of the $B$ band FP fully. 
Furthermore the derived dependence of $M/L_{K}$ on $L_{K}$ is obviously
inconsistent with that inferred from the $K$ band FP
(e.g., Djorgovski, Pahre, \& de Carvalho  1996).
This result suggests that the origin of the observed 
dependence of $M/L_{K}$ on $L_{K}$  cannot be  explained at all by the differences
in stellar populations between elliptical galaxies with different masses.

   The origin of the $K$ band FP has been already demonstrated not to be explained so simply
   by some effects of stellar population (Pahre \& Djorgovski 1997).
   The dependence of the mass-to-light ratio on mass inferred from the $K$ band FP
   is furthermore suggested to be explained 
   by structural and kinematical nonhomology in elliptical galaxies (Djorgovski  et al. 1996).
   Actually, numerical simulations  on dissipationless galaxy mergers (Capelato, de Carvalho
   \& Carlberg 1995) and dissipative ones (Bekki 1998) demonstrated that 
   structural and kinematical nonhomology is a natural result of violent relation
   combined with gaseous dissipation in galaxy merging.
   These observational and theoretical studies imply that the above failure in
   the successful reproduction of the $K$ band FP does not point out one
   of serious disadvantages of the present merger model but is on line with  
   the early suggestion that the $K$ band FP should be explained by the supposed mass-dependence
   of structure and kinematics (not by effects of stellar population). 
   Our future extensive dynamical studies of galaxy mergers will provide a plausible
   answer for the problem of whether or not the structural and kinematical nonhomology in merger remnants
   can  really reproduce the  $K$ band FP even if stellar populations of galaxy mergers 
   does not contribute at all to the creation of the $K$ band FP.

\placefigure{fig-11}
\placefigure{fig-12}
\placefigure{fig-13}

\subsection{Miscellaneous results}

 In the preceding sections, we have only
 addressed the importance of  initial disk mass ($M_{\rm d}$) 
 and supernova feedback ($f_{\rm sn}$) in determining a number of fundamental 
 $mass$-$dependent$  properties of elliptical galaxies.
 Here in \S 3.3, we emphasize the importance of parameters other than $M_{\rm d}$
 and $f_{\rm sn}$ in chemical evolution of gas-rich galaxy mergers.
 We particularly present the results of models with different chemical mixing length 
 $R_{\rm chem}$ (Model 10 and Model 11) and total amount of gaseous dissipation (Model 12 and 13).
 Both chemical mixing length and total amount of gaseous dissipation 
 are less critical for chemical evolution of galaxy mergers  than $M_{\rm d}$  and $f_{\rm sn}$. 
 It is, however, worth to investigate fundamental roles of these two parameters
 ($R_{\rm chem}$ and effectiveness of gaseous dissipation), since it has not been 
 observationally clarified yet
 what the most plausible values are for these two parameters.
 Furthermore we describe  mass-independent chemophotometric relations 
 derived  in merger remnants.

\subsubsection{Importance of other parameters}
 As our Paper I has already pointed out,
 other merger parameters, such as the mixing length of
 chemical components 
 and effectiveness of gaseous dissipation during galaxy merging,
 can greatly determine not only mean properties of stellar populations
 but also radial ones in merger remnants.
 The present study, which is more realistic and elaborated
 than the Paper I, should accordingly confirm the importance of
 other parameters of galaxy mergers.
 Figure 11 describes the dependence of the $U-R$ color gradient
 ($\Delta U-R / \Delta \log R$)
 on mean stellar metallicity ($\langle Z_{\ast} \rangle$) in merger remnants
 for six models with $M_{\rm d} = 10^{10} M_{\odot}$
 and $10^{12} M_{\odot}$ (Model 1, 2, 10, 11, 12, and 13).
 We include models with chemical mixing length  $R_{\rm chem}$ = 0.1 (Model 1 and 2), 
 those  with  $R_{\rm chem}$ = 0.4 (Model 10 and 11), 
 and those with  $R_{\rm chem}$ = 0.1 and  no gaseous dissipation (Model 12 and 13)
 in Figure 11,  and thereby observe how chemical mixing length 
($R_{\rm chem}$)
 and total amount of gaseous dissipation affect global and radial
 properties of stellar populations in merger remnants.
 The importance of chemical mixing length and gaseous dissipation
 clarified for $\langle Z_{\ast} \rangle$ and $\Delta U-R / \Delta \log R$
 can be applied to other mean and radial properties of metals,
 color, and line indices.
 As is shown in Figure 11, mean stellar metallicity is slightly smaller 
 in the models with larger $R_{\rm chem}$ (Compare, for example,
 the results of Model 1 with those of Model 10.),
 which  is consistent with that derived in the Paper I.
 The $U-R$ color gradient is not largely different between models with different 
 $R_{\rm chem}$ for less massive models whereas the 
 gradient is appreciably smaller in mergers with larger $R_{\rm chem}$
 for more massive models.
 This result implies that inhomogeneous chemical mixing described before (\S 3.1)
 is critical for the formation of color gradient especially for more massive galaxy
 mergers.
 Furthermore, mean stellar metallicity is found to be appreciably smaller  for models
 with no gaseous dissipation (Compare, for example, the results of Model 1 with
 those of Model 12.).
 This result clearly suggests that total amount of gaseous dissipation during
 galaxy merging is also important determinant for the mean properties of stellar
 populations of merger remnants.
 Figure 11 moreover shows that without gaseous dissipation,
 the $U-R$ color gradient in merger remnants becomes $positive$,
 which suggests that a sufficiently large amount of 
 gaseous dissipation is indispensable for successful reproduction
 of the observed negative color gradients of luminous elliptical galaxies.
 These result clearly demonstrate that both chemical mixing length
 and the degree of gaseous dissipation are important determinants
 for mean and radial chemical and photometric properties of merger remnants.
 These results accordingly imply that more careful numerical implementation
 on chemical mixing and gaseous dissipation in galaxy mergers is required
 for comparing more quantitatively numerical results with observational
 ones: The mass-dependent chemical, photometric, and spectroscopic
 properties of galaxy mergers clarified in the present study (especially in \S 3.1 and \S 3.2)
 can be meaningful only in a qualitative sense.

\subsubsection{Mass-independent chemophotometric  relations}

 We have so far restricted ourselves to mass-dependent 
 (or luminosity-dependent) chemical, photometric,
 and spectroscopic properties of elliptical galaxies.
 There are, however, a number of observational spectrophotometric properties
 that correlate not so well with galactic luminosity but strongly with
 other physical properties in elliptical galaxies.
 Figure 12 and 13 demonstrate two examples of this type
 of correlation  derived in the present study.
 Figure 12 shows the dependence of the radial gradient of $\rm Mg_{2}$ index 
 ($\Delta \log \rm Mg_{2} / \Delta \log \it R$) on
 the central $\rm Mg_{2}$ index (within  $0.1 R_{\rm e}$)
 in merger remnants for $all$ models in the present study.
 We can clearly observe that merger remnant with larger central
 $\rm Mg_{2}$ index has the larger (negative) gradient of $\rm Mg_{2}$ index.
 This result is consistent reasonably well with the recent observational
 result that the radial gradient of $\rm Mg_{2}$ index can  
 correlate with the central $\rm Mg_{2}$ index rather than with galactic
 luminosity or central velocity dispersion (Gonz\'alez \& Gorgas 1996).
 This result furthermore  shows that even if there is a great diversity 
 in parameters associated with physical processes of galaxy merging,
 the observed relationship between the central $\rm Mg_{2}$ index
 and $\Delta \rm Mg_{2} / \Delta \log \it R$ can be reproduced
 successfully by galaxy mergers:
 The origin of $\rm Mg_{2} - \Delta \rm Mg_{2} / \Delta \log \it R$
 relation does not depend on the details of galaxy merging processes. 
 Figure 13 gives the dependence  of stellar metallicity gradient 
 ($\Delta \log \langle Z \rangle / \Delta \log  R$) on the radial gradient 
 of the  epoch of star formation ($\Delta \log \langle T_{\rm SF} \rangle / \Delta \log  R$) 
 in merger remnants for $all$ models in the present study.
 The radial gradient of $\Delta \log \langle T_{\rm SF} \rangle / \Delta \log  R$
 corresponds to the age gradient in each merger remnant.
 As is shown in Figure 13,
 merger remnants with larger absolute magnitude of their age gradients
 have larger metallicity gradients.
 This result is consistent with the observational one inferred from
 the radial gradients of $\rm H_{\beta}$ and [MgFe] line indices
 (e.g., Chiosi 1996).
 The physical reason for the derived dependence is that
 irrespectively of merger parameters, younger stellar population
 in the central part of galaxy mergers are formed preferentially from
 more metal-enriched ISM transferred from the outer part of mergers
 owing to the efficient gas fueling. 
 Thus these two results on 
 the $\rm Mg_{2} - \Delta \rm Mg_{2} / \Delta \log \it R$ relation
 and $\Delta \log \langle T_{\rm SF} \rangle / \Delta \log  R -
 \Delta \log \langle Z \rangle / \Delta \log  R$ one provide clues to 
 the origins of observational results of Chiosi (1996) and 
 Gonz\'alez \& Gorgas (1996), which clearly show $mass$-$independent$  
 properties of elliptical galaxies.

\section{Discussion}

 We have  demonstrated that spectrophotometric properties of galaxy
 mergers depend on galactic mass (or luminosity),
 essentially because galactic mass can greatly affect
 both the rapidity of star formation and the effectiveness of supernova feedback
 in chemodynamical evolution of galaxies mergers.
 This implies that the origins of
 the observed fundamental characteristics of spectrophotometric
 properties of elliptical galaxies can be understood in terms of mass-dependent
 chemodynamical evolution of galaxy mergers.
 Here we particularly discuss whether or not the present merger model
 of elliptical galaxy formation can explain $quantitatively$ the origins of 
 the following two fundamental 
 empirical relationship between galactic mass (luminosity)
 and spectrophotometric properties 
 in elliptical galaxies: (1) The CM relation and (2) The $\rm Mg_{2}-\sigma$ one.

\subsection{The origin of the CM relation}

The CM relation is generally considered to be one
of important diagnostics which can confirm
whether or not a certain model of elliptical galaxy formation
is plausible and viable.
The present merger model predicts that more massive ellipticals
have larger stellar metallicity thus show redder colors.
The reasons for more efficient metal-enrichment in more massive
galaxy mergers are
firstly that  a smaller amount of metal-enriched ISM
can be tidally stripped away from galaxy mergers owing to
more rapid gas consumption by star formation,
secondly that star formation can be less strongly
suppressed by supernova feedback thus can produce a larger
amount of metals.
These results imply that the origin of the CM relation
of elliptical galaxies can be closely associated with
mass-dependent chemodynamical evolution of galaxy mergers.
Although these results can provide a
$qualitative$ explanation of the CM relation,
the present merger model still seems to have some  difficulties in explaining
$quantitatively$ the origin of the absolute magnitude of the
 CM slope (See Figure 8).
To be more specific,
the color difference in $V-K$ between a merger remnant with
$1.5 \times 10^{9} L_{\odot}$ and that with $1.9 \times 10^{11} L_{\odot}$  
in $B$ band
is only 0.45 (0.16) mag for models with $f_{\rm sn}=0.5$ (0.1), 
which seems  to disagree  with the observed
difference of 0.48 for the corresponding luminosity range
in Bower et al. (1992).
Since we did not  consider any age differences of galaxy mergers with different masses 
in the present study,
this $apparent$ disagreement implies that the present chemodynamical model
can not reproduce a   mass-metallicity relation,  which is one of possible interpretations  for 
the origin of the observed CM relation.
There are mainly two interpretations of the CM relation;
One  is that the CM relation is due only to a mass-metallicity relation,
i.e., the relation that more massive (luminous) ellipticals have larger metallicity
thus show redder colors,
and the other is that both a mass-metallicity relation and a mass-age relation,
i.e., the relation that more massive ellipticals are older,
play some roles in the formation of the CM relation.
We here reject the idea that only a systematical difference in mean galaxy ages between
ellipticals with different masses is a key determinant for the CM relation,
primarily because some theoretical studies have already demonstrated
that this idea is completely inconsistent with the observed redshift evolution
of the slopes of the CM relation (e.g., Kodama \& Arimoto 1997).
Accordingly,  whether or not the $apparent$ disagreement in the successful reproduction
of the CM relation is really a serious problem of the present merger model 
of elliptical galaxy formation
depends on which of the above two interpretations
we adopt.

Firstly,
we adopt 
the  conventional and classical  interpretation that 
elliptical galaxies $as$ $a$ $whole$ are formed at high redshift
and thus  there are not  any significant age differences 
between elliptical galaxies: The origin of the CM relation
can  be thus understood only in terms of
a mass-metallicity relation.
In this case,
the above apparent disagreement is really a serious problem 
of the present chemodynamical model of elliptical galaxy formation
and thus implies either that more sophisticated modification 
of the present chemodynamical model is necessary, including dissipative
dynamics, SNIa and SNII feedback effects, star formation parameters
(e.g, variable IMF), and chemical mixing processes,
or that typical luminous elliptical galaxies are not formed
by major disk-disk mergers.
Secondly,  we adopt the  interpretation that 
less luminous ellipticals are as a whole 
younger and less metal-enriched galaxies: The origin of the CM relation
is not necessarily a pure mass-metallicity relation,  and thus a systematic age difference between
ellipticals can also play a role in the formation of the CM relation
(e.g., Faber et al. 1995; Worthey et al. 1996).
In this case, the above apparent inconsistency can be resolved
by invoking the assumption that less luminous 
elliptical galaxies are formed by galaxy mergers with initially younger stellar populations
or by later
galaxy mergers.

  The considerably tight color-magnitude relation of elliptical galaxies 
  (Bower, Lucey, \& Ellis 1992; Ellis et al. 1997)
  and the  relatively smaller  redshift evolution  of photometric properties
  of elliptical galaxies 
  (Arag$\rm \acute{o}$n-Salamanca et al. 1993; Franx \& van Dokkum 1996)
  are generally considered to reflect the coeval formation of elliptical galaxies
  at high redshift.
  Furthermore the observed 
  considerably less significant evolution of
  the slope of the CM relation is suggested to reject 
  the age spread larger thatn 1 Gyr among elliptical galaxies
  (e.g., Kodama \& Arimoto 1997).
  These results imply that most of ellipticals are formed at high redshift
  and thus that the CM relation is a  pure mass-metallicity relation. 
  Luminosity-dependence of line index [MgFe] and $\rm H_{\beta}$ (Faber et al. 1995
  Worthey et al. 1996), 
  on the other hand, indicates that 
  less luminous ellipticals have younger stellar populations.
  These results imply that a systematic difference in mean galaxy ages between ellipticals
  can play some roles in the formation of the CM relation.
  Thus, it has not been observationally
  clarified  whether the observed CM relation is due essentially to   a pure 
  mass-metallicity relation or both a  mass-age relation   
  and  a mass-metallicity one 
  contribute to the CM relation.
  Considering these apparently inconsistent observational results, 
  it is reasonable for the present paper 
  to claim that whether or not the apparent failure of the present merger model  implies
  a serious disadvantage of the merger scenario of elliptical galaxy formation depends
  on future extensive observational studies on the mass-dependence of mean age and
  metallicity of elliptical galaxies.

\subsection{Chemodynamical evolution of mergers and the $\rm Mg_{2}$-$\sigma$ relation}

The present chemodynamical model predicts that
merger remnants with larger central velocity dispersion ($\sigma$)
have larger central $\rm Mg_{2}$ index,
principally because a larger amount of metal-enriched ISM 
can be transferred to the central region of galaxy mergers
and converted into stellar components there in more massive 
galaxy mergers with larger central velocity dispersion.
Although this result can give a qualitative explanation for 
the origin of $\rm Mg_{2}$-$\sigma$ relation of elliptical galaxies,
the present merger model has some difficulties in explaining
quantitatively the origin (See Figure 9).
For models with $f_{\rm sn}$ = 0.1,
a merger remnant with $\sigma= $ 305  km/s has $\rm Mg_{2}$ = 0.300 mag 
whereas that with  $\sigma=$ 105 km/s has $\rm Mg_{2}$ = 0.272 mag.
This result gives the relation $\rm Mg_{2} \propto 0.06 \log \sigma$.
On the other hand,
for models with $f_{\rm sn}$ = 0.5,
a merger remnant with $\sigma= $ 284  km/s has $\rm Mg_{2}$ = 0.291 mag
whereas that with  $\sigma=$ 47 km/s has $\rm Mg_{2}$ = 0.177 mag.
This result gives the relation ${\rm Mg}_{2} \propto 0.16 \log \sigma$.
The derived $\rm Mg_{2}$-$\sigma$
relations for the above  two cases ($f_{\rm sn}$ = 0.1 and 0.5)
are  not  consistent reasonably well with the observed relation 
${\rm Mg}_{2} \propto 0.2 \log \sigma$ (Burstein et al. 1988;
Davies 1996; J{\o}gensen 1997).
The reason  for this inconsistency is primarily that
the absolute magnitude of ${\rm Mg}_{2}$ for less luminous merger remnants
in the present study, in particular, for models with $f_{\rm sn}$ = 0.0 and 0.1,
is appreciably larger than the observed one.
We did  not consider any age differences of galaxy mergers with different masses 
in the present study.
Accordingly this inconsistency implies that the present chemodynamical model have  failed
to reproduce a  metallicity-$\sigma$ relation,  which is one of possible interpretations
for the origin of the observed
${\rm Mg}_{2}$-$\sigma$ relation.
The observed small scatter in the $\rm Mg_{2}$-$\sigma$ relation (Bender, Burstein, \& Faber 1993)
probably  allows the following two interpretations  on the origin of the $\rm Mg_{2}$-$\sigma$ relation.
One is that a   $\rm Mg_{2}$-$\sigma$ relation results exclusively  from
a  metallicity-$\sigma$ relation (i.e., the relation that a larger $\sigma$ galaxy
has larger stellar metallicity), and the other is that both a metallicity-$\sigma$ relation
and a  age-$\sigma$  (i.e., the relation that a larger $\sigma$ galaxy has older  stellar age) 
can contribute to the  $\rm Mg_{2}$-$\sigma$ relation
(We here consider that the observed small scatter in  the $\rm Mg_{2}$-$\sigma$ relation
rules out the interpretation that only a systematic age difference between ellipticals
can explain totally the origin of the $\rm Mg_{2}$-$\sigma$ relation).
Accordingly, whether or not our failure in the successful
reproduction of the $\rm Mg_{2}$-$\sigma$ relation is a serious problem of the present merger
model of elliptical galaxy formation
depends on which of the above two  interpretations  we adopt.

Firstly, we adopt the interpretation that the $\rm Mg_{2}$-$\sigma$ relation
reflects  only the dependence of stellar mean metallicity on $\sigma$ (or mass) in 
the central part of elliptical galaxies.
In this case, the $\rm Mg_{2}$-$\sigma$ relation derived in the present merger model
of elliptical galaxies is really inconsistent with observational one,
which implies  that
inward transfer of metals (Mg) during dissipative galaxy merging is not
so efficient as the present merger model predicts especially for
less luminous galaxy mergers.
For this case, we must either construct a more elaborated and realistic chemodynamical model
of galaxy mergers and thereby explore  again the origin of  the $\rm Mg_{2}$-$\sigma$ relation
or conclude that major disk-disk galaxy merging is not closely associated with the formation
of typical elliptical galaxies.
Secondly, we adopt the interpretation that less luminous ellipticals with smaller
$\sigma$ are more likely
to show  both smaller metallicity and  younger ages thus show smaller $\rm Mg_{2}$. 
In this case, 
our failure is not so serious problem for 
the present chemodynamical model of elliptical galaxy formation,
principally because we can reproduce the $\rm Mg_{2}$-$\sigma$ relation
by invoking the assumption that less luminous ellipticals with smaller
$\sigma$ are formed by mergers with initially younger stellar populations
or  by later mergers.

As is the case of the CM relation, 
it is observationally unclear which of the above two interpretations is more plausible and reasonable
for the origin of the $\rm Mg_{2}$-$\sigma$ relation.
Observation studies on the redshift evolution
of the $\rm Mg_{2}$-$\sigma$ relation in cluster ellipticals (Ziegler \& Bender 1997),
the color-magnitude diagram of  stellar populations of  M32  by $HST$ (Grillmair et al. 1996),
and the dependence of the slope of the FP on the observational wavelength (Pahre et al. 1998) 
indicate younger stellar populations in less luminous ellipticals.
The considerably smaller scatter in the  $\rm Mg_{2}$-$\sigma$ relation, however,
implies coeval formation of elliptical galaxies at higher redshift universe (Bender et al. 1993).
Furthermore there are  no observational studies, at least now, which
clearly demonstrate  that less luminous ellipticals $as$ $a$ $whole$
have younger stellar populations in the central part of galaxies.
We are accordingly
conservative to determine the more plausible interpretation for the $\rm Mg_{2}$-$\sigma$ relation.  
Thus, it is safe for us to say that
it depends on
more extensive future observational studies
whether or not the present numerical results
on the $\rm Mg_{2}$-$\sigma$ relation 
are really consistent with observational ones.

\section{Conclusion}

We  have numerically investigated  chemodynamical 
evolution of major disk-disk galaxy mergers 
in order to explore the origin of mass-dependent chemical, 
photometric, and spectroscopic properties observed in  elliptical galaxies.
The present study is an extended version of our previous
study (Paper I),
which is the first step toward the understanding 
of a close physical relationship
between dynamical evolution and chemical one
in galaxy mergers.
 The present chemodynamical model
 is more realistic and sophisticated than our Paper I 
 in the following three points:
 (1) Instead using instantaneous chemical mixing approximation,
 we consider time-delay between star formation
 and metal ejection mainly from Type Ia and II supernovae.
 (2) Feedback effects of Type Ia and II supernovae 
 on dynamical and chemical evolution of galaxy mergers
 are incorporated.
 (3) Time-evolution of  enrichment processes 
 is  solved for each of chemical components, 
 H, He, Mg, O, and Fe.
We have found that initial galactic mass ($M_{\rm d}$)
is one of critical determinants  for chemical,
photometric, and spectroscopic properties of merger remnants.
The essential  reason for the derived mass-dependence
is that galactic mass 
can largely determine
total amount of metal-enriched interstellar gas,
star formation histories of galaxy mergers, and the effectiveness of Type Ia and II  
supernova feedback,
all of which greatly affect chemodynamical evolution of galaxy mergers.
In particular, the difference in the effectiveness of Type Ia and II
supernova feedback between galaxy mergers with different masses
is found to determine not only the $mean$ properties but also
the $radial$ ones in ellipticals with different masses.

      What we should emphasize here is that since this paper is only the first step
      toward the deep understanding of chemodynamical and spectrophotometric evolution
      of galaxy mergers, we completely neglect some possibly important effects of galactic
      dynamics on chemophotometric evolution of galaxies.
      Accordingly, some dynamical effects that are not investigated in the present paper
      can greatly modify the derived mass-dependence of chemophotometric properties
      of merger remnants. Among these,  dynamical effects of galactic bulges are
      particularly important in  the sense that bulges can largely control 
      inward and radial  mass-transfer
      of interstellar gas (Mihos \& Hernquist 1996) and consequently determine 
      star formation histories  of galaxy mergers and the nature of stellar populations formed by
      the induced secondary starbursts.
      Probably these dynamical effects of bulges  modify  the absolute magnitude of
      radial gradients of spectrophotometric properties derived in the present 
      merger model without bulges. 
      Our future studies with a more realistic merger model
      will investigate in detail how galaxy bulges affect the nature of stellar populations
      of merger remnants.

Main results obtained in this study are the following seven.

(1) More massive (luminous) ellipticals formed by galaxy mergers between more
massive spirals have larger metallicity and thus show redder colors,
principally because a larger amount of metal-enriched ISM can be converted
into stellar components in more massive galaxy mergers.
This result implies that the origin of the CM relation of elliptical
galaxies can be closely associated with dissipative galaxy merging
with star formation.
The typical metallicity ranges
from $\sim$ 1.0 solar abundance ($Z \sim 0.02$) for ellipticals
formed by mergers with $M_{\rm d} =  10^{10} M_{\odot}$
to $\sim$ 2.0 solar (Z $\sim 0.04$) for those with $M_{\rm d} =  10^{12} M_{\odot}$.

(2) The absolute magnitude of negative metallicity gradients developed in
galaxy mergers  is more likely to be larger for massive ellipticals.
Typical value of metallicity gradient 
($\Delta \log \langle Z_{\ast} \rangle / \Delta \log R$) for $0.1 \leq R/R_{\rm e} \leq 1.0$ 
 (where $R_{\rm e}$ is  effective radius of a merger remnant)
 is -0.13 for ellipticals with $M_{\rm d} =  10^{12} M_{\odot}$
 and -0.03 for those with $M_{\rm d} =  10^{10} M_{\odot}$.
 Absolute magnitude of metallicity gradient  correlates  with that of
 age gradient in  ellipticals
 in the sense that an elliptical  with steeper negative metallicity gradient
 is more likely to show
 steeper age gradient.

(3) Irrespective of galactic mass, a stellar population in the central part 
is both younger and more metal-enriched than that 
in the outer part  in an elliptical galaxy formed by dissipative galaxy
merging with star formation.
This is essentially because metal-enriched ISM
can be  transferred to the central part of galaxy mergers
owing to efficient gas fueling during dissipative galaxy merging with star
formation and preferentially
converted into stellar components there in the later phase of the merging.

(4) Abundance ratio of [Mg/Fe] does not depend so strongly on galactic mass 
(typically $0.2 \sim 0.3$ solar).
The absolute magnitude of the radial gradient of [Mg/Fe] is rather small
and does not depend  on galactic mass ($M_{\rm d}$).

(5) Radial color gradient is  more likely to
be larger for more massive ellipticals,
which reflects that the metallicity gradient is larger for more massive ellipticals.
For example,
typical $U-R$ color gradient  
 ($\Delta U-R / \Delta \log R$)
 for  $0.1 \leq R/R_{\rm e} \leq 1.0$
 is -0.13 for ellipticals with $M_{\rm d} =  10^{12} M_{\odot}$
 and -0.07 for those with $M_{\rm d} =  10^{10} M_{\odot}$.

(6) $\rm Mg_{2}$ line index in the 
central part of ellipticals ($R \leq 0.1 R_{\rm e}$) 
is larger for more massive ellipticals,
principally because a larger amount of metal-enriched ISM
can be fueled to the central part of mergers and converted into
stellar components there  in more massive galaxy mergers.
This result implies that the origin of $\rm Mg_{2}-\sigma$ relation observed
in elliptical galaxies can be closely associated with the difference
in total amount of metal-enriched ISM transferred to galactic center 
between galaxy mergers with different masses.
The radial gradient of $\rm Mg_{2}$ ($\Delta \rm Mg_{2} / \Delta \log R$) 
is also   more likely to be larger for massive ellipticals. 
$\Delta \rm Mg_{2} / \Delta \log R$  correlates well  with
the central $\rm Mg_{2}$ in ellipticals.
For most of the present merger models,
ellipticals show $positive$ radial gradient of $\rm H_{\beta}$ line index.

(7) Both $M/L_{B}$ and $M/L_{K}$, where $M$, $L_{B}$, and $L_{K}$
are total stellar mass of galaxy mergers, $B$-band luminosity, and $K$-band
one, respectively, depend on galactic luminosity  
in such a way that more luminous  ellipticals have
larger $M/L_{B}$ and smaller $M/L_{K}$.
This result reflects that more luminous  merger remnants are more likely to
have larger mean stellar metallicity.

These seven numerical results derived in the
present chemodynamical merger model of elliptical galaxy
formation imply that the observed fundamental dependence of
chemophotometric properties of ellipticals
on galactic mass or luminosity (e.g., the CM relation,
the $\rm Mg_{2}-\sigma$ relation, and the FP)
can be understood in terms of mass-dependent chemodynamical
evolution of galaxy merges.

\acknowledgments

K.B. thanks to the Japan Society for Promotion of Science (JSPS) 
Research Fellowships for Young Scientist.

\newpage

\appendix

\section{Chemical, photometric, and spectroscopic
evolution of SSP with variously  different IMF
and $M_{\rm low}$}

Chemical, photometric, and spectroscopic evolution of galaxies are generally
considered to depend not only on
star formation histories of galaxies resulting from dynamical evolution of galaxies,
as is investigated  in the present study, but also on detailed properties of IMF,
$M_{\rm low}$, and  $M_{\rm up}$.
Accordingly we should emphasize that the results derived in the present study (with
the Salpeter IMF, $M_{\rm low}=0.6 M_{\odot}$,
and $M_{\rm up}=120.0 M_{\odot}$) are {\it only true for} galaxy mergers with
the Salpeter IMF, $M_{\rm low}=0.6 M_{\odot}$, and $M_{\rm up}=120.0 M_{\odot}$
in a quantitative sense: We should admit limited applicability of the present 
chemodynamical model.
We here present dependence of chemical, photometric, and spectroscopic evolution
of a singe stellar population (SSP) on  IMF, $M_{\rm low}$, and  $M_{\rm up}$,
and thereby show more clearly 
how strongly or weakly the results derived in the present numerical results can
depend on IMF, $M_{\rm low}$,  and  $M_{\rm up}$.
In particular, we describe (1) Time evolution of ejected metals in a SSP 
with variously different IMF and  $M_{\rm low}$,
and (2) Photometric and spectroscopic 
evolution of the SSP.
These two descriptions  
will help us to understand how numerical results derived in the present
study can be changed if we adopt IMF, $M_{\rm low}$, and  $M_{\rm up}$ other
than those adopted in the present study.

\subsection{Time evolution of stellar yield ($Y_{i,j}(t)$)}

Time evolution of total amount of metals (($Y_{i,j}(t)$) ejected from stars 
with variously deferent mass and lifetime in a SSP with
a given metallicity depends on 
IMF and  $M_{\rm low}$.
We here observe how the time evolution of O, Fe, Mg, and Z
ejected from stars depends on IMF and $M_{\rm low}$,
by showing the  results for  models with
the Salpeter IMF, $M_{\rm low}$=0.1$M_{\odot}$, 
and $M_{\rm up}$=120.0$M_{\odot}$ (Model A),
with the Salpeter IMF,  $M_{\rm low}$=0.6$M_{\odot}$,
and $M_{\rm up}$=120.0$M_{\odot}$ (Model B),
and with  IMF with shallower slope (the exponent of IMF equal to
$-1.1$), $M_{\rm low}$=0.1$M_{\odot}$, and $M_{\rm up}$=120.0$M_{\odot}$ (Model C).
The time evolution of Model B corresponds to $Y_{i,j}(t)$
adopted in the main manuscript.
It is clear from Figure 14 that (1) Z is larger in a SSP with larger $M_{\rm low}$
for a given  IMF, (2) Z is larger in a SSP with shallower IMF slope
for a given $M_{\rm low}$, and (3) The results of (1) and (2)  can be applied
to O, Mg, and Fe.
These three results, though based on SSP analysis,
are well consistent with previous studies solving fully  
chemical evolution of closed-box one-zone galactic models 
(Arimoto \& Yoshii 1987; Vazdekis et al. 1996). 
These results accordingly imply that the {\it absolute magnitude} 
of present numerical results are changed 
if we adopt IMF and $M_{\rm low}$ other than 
those adopted in the present study.
For example, if we adopt smaller $M_{\rm low}$ (shallower IMF slope),
mean stellar metallicity of galaxy mergers can become 
substantially smaller (larger).
Thus, we again stress that although {\it the relative importance} 
of each parameter (e.g., galactic mass) 
in chemophotometric evolution of galaxy mergers derived
in the present study does not depend on 
IMF and $M_{\rm low}$, the derived {\it absolute magnitude} 
of chemophotometric properties in merger remnants should be 
more carefully interpreted owing
to dependences of chemical properties on IMF and $M_{\rm low}$  
derived in the above three results.

 In order to assess the validity of $Y_{i,j}(t)$ adopted in the present study,
 we furthermore check whether or not the  
 time-evolution of stellar yield ($Y_{i,j}(t)$) adopted in
 the present study can reproduce reasonably well  even 
 the chemical evolution of  the abundance ratio of [Mg/Fe], which has been already investigated
 by classical one-zone chemical evolution models (e.g., Matteucci \& Tornamb\`{e} 1987).
 We investigate the $long$-$term$ evolution ($\sim$ 10 Gyr)
 of stellar [Mg/Fe]  
 (the ratio of  stellar Mg  abundance to stellar Fe one in units of solar [Mg/Fe]) 
 produced by stellar systems with three different types
 of SSP in Model A, B, and C.
 We check whether or not the [Mg/Fe] can finally be 0.0 that is a typical value 
 for the solar neighborhood in the Galaxy, in which chemical evolution is considered to 
 proceed for more than 10 Gyr. 
 Figure  15 describes the time evolution of [Mg/Fe] for Model A, B, and C.
 It is clear from Figure 15 that (1) Model B, the $Y_{i,j}(t)$ of which
 is exactly the same as that used in the main manuscript,
 can reproduce the typical value of [Mg/Fe] 
 in the solar neighborhood in the Galaxy ([Mg/Fe]$\sim$0.0) after long-term evolution,
 and  (2) Final value of [Mg/Fe] does not depend so strongly on IMF and $M_{\rm low}$.
 These  results accordingly justify the $Y_{i,j}(t)$ adopted in the present study.
 Considering the [Mg/Fe] value of 0.2$\sim$0.3 derived for galaxy mergers
 and that of $\sim$ 0.0 for the above SSP models with long-term and less rapid
 chemical evolution,
 the above  two results
 furthermore imply  that the higher values of  [Mg/Fe] derived in the present
 merger model is closely associated with dissipative galaxy merging with
 rapid star formation.

\placefigure{fig-14}
\placefigure{fig-15}
\placefigure{fig-16}

\subsection{Time evolution of spectrophotometric properties in SSP}

 We  show how fundamental characteristics of spectrophotometric
 evolution of SSP  with solar metallicity 
 depend on IMF and $M_{\rm low}$ 
 in Figure  16 for Model A, B, and C.
 In calculating spectrophotometric properties of SSP,
 we adopt  the PEGASE  code of Fioc \& Rocca-Volmerange (1997) 
 which can reproduce the GISSEL96 SSP. 
 It is clear from Figure 16 that (1) The ratio of stellar mass ($M$)
 to luminosity ($L_{\rm B}$) is strongly dependent on IMF and $M_{\rm low}$,
 and (2) The $B-V$ and $V-K$ colors are not so different between models
 with different IMF and $M_{\rm low}$.
 In the present study, we derive spectrophotometric properties
 of merger remnant by using the GISSEL96 SSP  in
 which the Salpeter IMF and $M_{\rm low}=0.1 M_{\odot}$
 are assumed,
 nevertheless we adopt the Salpeter IMF and $M_{\rm low}=0.6 M_{\odot}$
 (That is, we  adopt a certain `approximation' in deriving the
 spectrophotometric properties.).
 Accordingly, spectrophotometric evolution of galaxy mergers
 is not self-consistently solved in a strict sense
 and more careful interpretations for results derived in the present
 study are necessary.
 Observing  the above two results on the differences in
 spectrophotometric properties of SSP between Model A $\sim$ C,
 we can assess the validity of the `approximation':
 Global integrated colors derived
 in the present study can be approximately correct
 whereas mass-to-light-ratio can $not$. 
 Thus we must again emphasize that discussing the dependence
 of spectrophotometric properties of merger remnants
 on galactic parameters such as initial disk mass
is considerably sensible whereas comparing the {\it absolute magnitude}
 of the present numerical results (in particular,  mass-to-light-ratio)
 with those of observational ones
 is not plausible: What is more vital in the present study
 is to clarify the $relative$  $importance$  of galactic parameters (e.g.,
 galactic mass) in determining spectrophotometric properties of merger remnants.

\section{A possible luminosity dependence of $C_{\rm SF}$ and $f_{\rm eff}$}

The essential reason for mass-dependent properties of galaxy mergers
is that galactic mass can determine both $C_{\rm SF}$ and $f_{\rm eff}$.
We here describe the reason for the mass-dependence of $C_{\rm SF}$
and $f_{\rm eff}$.
Observational background for the  mass-dependence of $C_{\rm SF}$
is given in detail in our Paper I.
Firstly  the expected mass dependence of  $C_{\rm SF}$ is described 
as follows.
The parameter $C_{\rm SF}$ is set to be proportional to 
 $T_{\rm dyn}$/$T_{\rm SF}$,
where  $T_{\rm dyn}$ and $T_{\rm SF}$  are the  dynamical time-scale
and the time-scale of gas consumption by star formation,
respectively.
We here define  the mass of a galactic disk, the total mass   of 
luminous  and dark matter,  and size of the progenitor as 
$M_d$, $M_t$, and $R_d$, respectively.
We consider here 
that gas mass in a disk is equal to $M_d$ for simplicity. 
The $T_{\rm dyn}$ is given as  
\begin{equation}
T_{\rm dyn} \propto R_d^{3/2} M_t^{-1/2}
\end{equation}
Provided that the coefficient in the Schmidt law is not dependent on the
galactic mass (or luminosity), we can derive $T_{\rm SF}$ as follows. 
\begin{equation}
T_{\rm SF} \propto \Sigma^{1 - \gamma} \; ,
\end{equation}
where $\Sigma$ is the surface density of the gas disk. 
The parameter $\gamma$ is the exponent of Schmidt law, which is the same as 
that used in previous subsections. 
Assuming the Freeman's law and the constant ratio of $R_d$ to 
the scale length of exponential disk, 
we derive
\begin{equation}
\Sigma \propto M_d R_d^{-2} \sim {\rm const.} 
\end{equation}
Assuming that the degree of self-gravity of a galactic disk is described as 
\begin{equation}
M_t \propto M_d^{(1 - \beta)} \; \; ,
\end{equation} 
then we can derive 
\begin{equation}
C_{\rm SF} \propto M_d^{1/4 + \beta/2} \; \; .
\end{equation}
Since $\beta$ is considered to have positive value 
($\beta = 0.6$: Saglia 1996), 
this relationship predicts that $C_{\rm SF}$ becomes larger as 
$M_d$ increases.
If we adopt the observed trend that more luminous disks have larger  surface 
density, such as $\Sigma \propto M_d$ (McGaugh \& Blok 1997), instead of using
the Freeman's law, we can obtain
$C_{\rm SF} \propto M_d^{1/2 + \beta/2}$.
Since in the present study, we assume, for simplicity, that the ratio of dynamical mass
to initial disk mass is constant for galaxy mergers with different masses (that is,
$\beta = 0.0$), and furthermore that  $M_{\rm d}$ is proportional to $R_d^{2}$ (Freeman's law),
the $M_{\rm d}$ dependence of $C_{\rm SF}$ adopted in the present study is
described as:
\begin{equation}
C_{\rm SF} \propto M_d^{0.25} \; \; .
\end{equation}

Secondly the expected mass dependence of  $f_{\rm eff}$ is described 
as follows.
If we neglect the importance of details of 
three dimensional structure of a  galaxy with dark matter halo
in determining the total potential energy of the galaxy,
total potential energy of an  initial disk ($W_{\rm d}$) is then considered to be
within an order of 
$GM_{\rm t}^{2}/R_{\rm d}$.
Assuming that the total  number of supernovae
is  basically dependent on $M_{\rm d}$,
the total energy of supernovae that can be potentially accumulated
within  the disk ($E_{\rm SN}$)
is proportional to $e_{\rm sn} M_{\rm d}$, where $e_{\rm sn}$ is total energy
produced by $one$ supernova ($\sim 10^{51}$ erg both for SNIa and SNII).
Accordingly, the ratio of total supernova energy (that can be ejected
in the course of galactic evolution) to total potential energy of the galaxy,
which corresponds to the effectiveness of supernova feedback in galactic
evolution,
is as follows:
\begin{equation}
f_{\rm eff} \propto E_{\rm SN}/W_{d} \propto
e_{\rm sn} M_{\rm d}/GM_{\rm t}^{2}/R_{\rm d} \propto R_{\rm d}/M_{\rm d}
\propto M_{\rm d}^{-0.5}  \; \; .
\end{equation}
In the above equation, the constant $M_{\rm t}/M_{\rm d}$ and  
 $M_{\rm d} \propto  R_d^{2}$ are assumed.
 This equation means that supernova feedback effects can not so strongly
 affect  evolution of galaxies for more massive galaxy mergers. 
 It should be emphasized here that the expected $M_{\rm d}$ dependences
 of $C_{\rm SF}$ and  $f_{\rm eff}$ are based on some assumptions 
 adopted for deriving these dependence.

\clearpage






\figcaption{Time evolution of gas mass fraction ($f_{\rm g}$)
and mean stellar metallicity ($\langle Z_{\ast} \rangle$) for models
with $M_{\rm d} = 10^{10} M_{\odot}$ (Model 1, solid lines)
and $M_{\rm d} = 10^{12} M_{\odot}$ (Model 2, dotted ones).
For the two models, the time evolution for 16 dynamical time ($t_{\rm dyn}$)
of each galaxy merger is described.
Note that both stellar mass fraction and mean stellar metallicity
are  larger for the more massive galaxy merger.
\label{fig-1}}

\figcaption{Final metallicity distribution of stellar component ($Z_{\ast}$) in
merger remnants at $T=16 t_{\rm dyn}$ for models with $M_{\rm d} = 10^{10} M_{\odot}$ 
(Model 1, shading bars) and $M_{\rm d} = 10^{12} M_{\odot}$ (Model 2, open ones).
The time $T=16 t_{\rm dyn}$ corresponds to $\sim$ 1.4 Gyr for Model 1
and $\sim$ 4.6 Gyr for Model 2.
Fractional mass of stellar components with a given metallicity (in $\log Z_{\ast}$)
is shown for each metallicity bin.
Note that the  more massive galaxy merger has a larger amount of substantially
metal-enriched stellar populations.
\label{fig-2}}

\figcaption{Time evolution of mean stellar metallicity of
$\langle Z_{\ast, Fe} \rangle$ (top) and $\langle Z_{\ast, Mg} \rangle$ (middle) and that of the
abundance ratio of
[Mg/Fe] (bottom) in merger remnants 
for models with $M_{\rm d} = 10^{10} M_{\odot}$ (Model 1, solid lines)
and $M_{\rm d} = 10^{12} M_{\odot}$ (Model 2, dotted ones).
For the two models, the time evolution for 16 dynamical time
of each galaxy merger is described.
\label{fig-3}}

\figcaption{
Radial profiles of metallicity gradients for 
$\langle Z_{\ast}  \rangle$ (top),  $\langle Z_{\ast, Fe} \rangle $ (second from the top), 
and $\langle Z_{\ast, Mg} \rangle $ (second from the bottom) and that of
[Mg/Fe] in merger remnants at $T=16 t_{\rm dyn}$
for models with $M_{\rm d} = 10^{10} M_{\odot}$  (Model 1, solid lines)
and with $M_{\rm d} = 10^{12} M_{\odot}$  (Model 2, dotted ones).
The time $T=16 t_{\rm dyn}$ corresponds to $\sim$ 1.4 Gyr for Model 1
and $\sim$ 4.6 Gyr for Model 2.
\label{fig-4}}

\figcaption{ 
Radial profiles of mean epoch of star formation 
($\langle T_{\rm SF} \rangle$) in merger remnants at $T=16 t_{\rm dyn}$
for models with $M_{\rm d} = 10^{10} M_{\odot}$  (Model 1, solid lines)
and with $M_{\rm d} = 10^{12} M_{\odot}$  (Model 2, dotted ones).
The larger value of $\langle T_{\rm SF} \rangle$ means younger stellar populations
in each radius.
The time $T=16 t_{\rm dyn}$ corresponds to $\sim$ 1.4 Gyr for Model 1
and $\sim$ 4.6 Gyr for Model 2.
Note that irrespectively of mass of galaxy mergers,
stellar populations in the central part of merger remnants are
younger than those in the outer part of merger remnants.
\label{fig-5}}

\figcaption{ 
Radial profiles of color gradients for 
$U-R$ (top),  $B-R$ (middle) 
and $V-K$  (bottom) 
in merger remnants at the age of 13.1 Gyr 
for models with $M_{\rm d} = 10^{10} M_{\odot}$  (Model 1, solid lines)
and with $M_{\rm d} = 10^{12} M_{\odot}$  (Model 2, dotted ones).
The rapid increase and decrease of  colors within $0.1 R_{\rm e}$
are due to the  smaller particle number in the central part of  each merger remnant. 
For comparison, the observed  mean  color gradients (Peletier et al. 1990a, b) 
with  arbitrary zero-points
are  given by  dash-dotted  lines. 
Here mean values of color gradients for $U-R$ (top),   $B-R$ (middle), and $V-K$  (bottom)
are -0.20,   -0.09, and -0.14,  respectively.
\label{fig-6}}

\figcaption{
Radial profiles of line indices for 
$\rm H_{\beta}$ (top),  $\rm Mg_{2}$ (middle) 
and $\langle Fe \rangle$ (bottom) 
in merger remnants at the age of 13.1 Gyr 
for models with $M_{\rm d} = 10^{10} M_{\odot}$  (Model 1, solid lines)
and with $M_{\rm d} = 10^{12} M_{\odot}$  (Model 2, dotted ones).
The rapid increase and decrease of  line indices  within $0.1 R_{\rm e}$
are  due to the  smaller particle number in the central part of each merger remnant. 
For comparison, the observed  mean  $\rm Mg_{2}$  gradient (Davies, Sadler,  \& Peletier 1993)
with  arbitrary zero-points
are given by  dash-dotted  lines. 
Here mean values of line index  gradients for $\rm H_{\beta}$  (top),    $\rm Mg_{2}$  (middle),
and $\langle Fe \rangle$ (bottom) 
are 0.23, -0.059, and  -0.38, respectively.
\label{fig-7}}

\figcaption{
The luminosity ($L_{B}$ in units of $L_{\odot}$) dependence of 
$U-R$ (top),  $B-R$ (middle), 
and $V-K$  color (bottom)  
in merger remnants at the age of 13.1 Gyr,
which corresponds to the color-magnitute (CM) relation
of elliptical galaxies in the present universe,
for models with different masses 
($M_{\rm d}$ = $10^{10}$, $10^{11}$, and $10^{12} M_{\odot}$) 
and strength of supernova
feedback effects ($f_{\rm sn}$ = 0.0, 0.1, and 0.5).
The results for models with $f_{\rm sn}$ = 0.0 (no feedback effects),
0.1, and 0.5 are plotted by open squares, circles, and triangles,
respectively.
The results with larger $L_{B}$ mean those with larger $M_{\rm d}$.
For clarity, the results for models with the same $f_{\rm sn}$
are connected by a dotted  line.
For comparison, the observed  $V-K$ color-magnitude relation (Bower et al. 1992) 
is given by a dash-dotted  line in the bottom   panel. 
Note that the CM relation can be more discernably observed
in models with larger $f_{\rm sn}$.
\label{fig-8}}

\figcaption{
The dependence of the central $\rm Mg_{2}$ index (within  $0.1R_{\rm e}$)
on the central velocity dispersion ($\sigma$) in merger remnants
at the age of 13.1 Gyr
for models with different masses 
($M_{\rm d}$ = $10^{10}$, $10^{11}$, and $10^{12} M_{\odot}$) 
and strength of supernova
feedback effects ($f_{\rm sn}$ = 0.0, 0.1, and 0.5).
The results for models with $f_{\rm sn}$ = 0.0 (no feedback effects),
0.1, and 0.5 are plotted by open squares, circles, and triangles,
respectively.
The results with larger $\sigma$  mean those with larger $M_{\rm d}$.
For clarity, the results for models with the same $f_{\rm sn}$
are connected by a dotted  line.
For comparison, the observed  $\rm Mg_{2}-\sigma$ relation  (Burstein et al. 1988) 
is given by a dash-dotted  line. 
Note that the  $\rm Mg_{2}-\sigma$ relation can be more discernably observed
in models with larger $f_{\rm sn}$.
\label{fig-9}}

\figcaption{
The luminosity ($L_{B}$ and $L_{K}$ in units of $L_{\odot}$) dependence of 
mass-to-light-ratio ($M/L_{B}$ and $M/L_{K}$)
in merger remnants at the age of 13.1 Gyr
for models with different masses 
($M_{\rm d}$ = $10^{10}$, $10^{11}$, and $10^{12} M_{\odot}$) 
and strength of supernova
feedback effects ($f_{\rm sn}$ = 0.0, 0.1, and 0.5).
Here mas-to-light-ratio means the ratio of total $stellar$ mass (not total
mass of a merger remnant) to total stellar
light in a merger remnant. 
The results for models with $f_{\rm sn}$ = 0.0 (no feedback effects),
0.1, and 0.5 are plotted by open squares, circles, and triangles,
respectively.
The results with larger $L_{B}$ and $L_{K}$ mean those with larger $M_{\rm d}$.
For clarity, the results for models with the same $f_{\rm sn}$
are connected by a dotted  line.
For comparison, the observed luminosity dependeces of $M/L_{B}$  
($M/L_{B} \propto {L_{B}}^{0.2}$ derived by  Bender et al. 1992)
and $M/L_{K}$ ($M/L_{K} \propto {L_{K}}^{0.17}$ by Pahre et al. 1998)
are also given by solid lines with arbitrary zero-points.
\label{fig-10}}

\figcaption{The dependence of $U-R$ color gradient 
on mean stellar metallicity ($\langle Z_{\ast} \rangle$) in merger
remnants for models with $R_{\rm chem}=0.1$ and gaseous dissipation
(Model 1 and 2, open circles),
those with $R_{\rm chem}=0.4$ and gaseous dissipation
(Model 10 and 11, filled squares),
and those with $R_{\rm chem}=0.1$ and $no$ gaseous dissipation
(Model 12 and 13, filled circles).
In this figure, the results for three models with  
$M_{\rm d} = 10^{10} M_{\odot}$ (Model 1, 10, and 12)
and those for three models with $M_{\rm d} = 10^{12} M_{\odot}$ (Model 2, 11, and 13) 
are given.
The results with larger $\langle Z_{\ast} \rangle$ denote those for models with 
$M_{\rm d} = 10^{12} M_{\odot}$.
For clarity, the results for models with the same $R_{\rm chem}$ and degree of 
gaseous dissipation (for example, Model 1 and 2) are connected by a dotted line. 
\label{fig-11}}

\figcaption{
 The dependence of  the radial gradient of $\rm Mg_{2}$ index
 ($\Delta  \rm Mg_{2} / \Delta \log \it R$)
 on the central $\rm Mg_{2}$ index in merger remnants for all models
 in the present study. For comparison,  the observed depdence of
 Gonz\'alez \& Gorgas (1996),
 $\Delta  \rm Mg_{2} / \Delta \log  R  = 0.026-0.28\rm Mg_{2}$,
 is also given by a solid line. 
\label{fig-12}}

\figcaption{
The dependence of the radial gradient of stellar metallicity  
 ($\Delta \log \langle Z_{\ast} \rangle / \Delta \log  R$) on that of mean epoch
 of star formation  
($\Delta \log \langle T_{\rm SF} \rangle / \Delta \log  R$)
 in merger remnants for all models
 in the present study.
 Here merger remnants with  larger value of $\langle T_{\rm SF} \rangle$ 
 have a larger amount of younger stellar populations.
 Note that merger remnants with steeper age gradient
 show steeper metallcity gradient.
\label{fig-13}}

\figcaption{
Time evolution of total mass in units of $M_{\odot}$
for  metals Z ($m_{\rm Z}$, top), O ($m_{\rm O}$, second from the top),
Fe ($m_{\rm Fe}$, second from the bottom), and Mg ($m_{\rm Mg}$, bottom),
ejected from stars with variously different masses and lifetimes
in a SSP with solar metallicity and solar mass ($1.0M_{\odot}$)
for the following three models:
Model A with the Salpeter IMF, $M_{\rm low}$=0.1$M_{\odot}$, 
and $M_{\rm up}$=120.0$M_{\odot}$ (solid lines),
Model B  with the Salpeter IMF,  $M_{\rm low}$=0.6$M_{\odot}$,
and $M_{\rm up}$=120.0$M_{\odot}$ (dotted lines),
and Model C  with  IMF with shallower slope (the exponent of IMF equal to
-1.1), $M_{\rm low}$=0.1$M_{\odot}$, and $M_{\rm up}$=120.0$M_{\odot}$ (dashed lines).
The time evolution of  Model B corresponds to $Y_{i,j}(t)$
adopted in the main manuscript.
\label{fig-14}}

\figcaption{
Long-term  evolution of  the abundance ratio of [Mg/Fe] in
a one-zone chemical evolution model of a disk galaxy for 
Model A (solid line), B (dotted one), and C (dashed one).
Details of the one-zone disk model are given in the manuscript.
\label{fig-15}}

\figcaption{Time evolution of $M/L_{B}$, $B-V$, and $V-K$ color
in a SSP with solar metallicity for Model A (solid lines), B (dotted ones), and C (dashed ones).
Here $M/L_{B}$ means the ratio of stellar mass to $B$-band luminosity.
Note that $M/L_{B}$ depends more strongly on IMF and $M_{\rm low}$
whereas $B-V$ and $V-K$ color do not.
\label{fig-16}}

\end{document}